\documentclass[aip,jcp,numerical,reprint]{revtex4-1}

\usepackage{float}
\restylefloat{table}
\usepackage{amsmath}
\usepackage{graphicx}
\usepackage{dcolumn}
\usepackage{bm}
\usepackage{braket}
\usepackage{gensymb}
\usepackage{array}
\newcolumntype{P}[1]{>{\centering\arraybackslash}p{#1}}
\newcolumntype{M}[1]{>{\centering\arraybackslash}m{#1}}

\begin{document}

\title{Ultra-Fast Relaxation, Decoherence and Localization of Photoexcited States in $\pi$-Conjugated Polymers: A TEBD Study}

\author{Jonathan R. Mannouch}\affiliation{Department of Chemistry, Physical and Theoretical Chemistry Laboratory, University of Oxford, Oxford, OX1 3QZ, United Kingdom}\affiliation{University College, University of Oxford, Oxford, OX1 4BH, United Kingdom}
\author{William Barford}\affiliation{Department of Chemistry, Physical and Theoretical Chemistry Laboratory, University of Oxford, Oxford, OX1 3QZ, United Kingdom}
\author{Sarah Al-Assam}\affiliation{Department of Physics, Clarendon Laboratory, University of Oxford, Oxford, OX1 3PU, United Kingdom}

\begin{abstract}
The exciton relaxation dynamics of photoexcited electronic states in poly($p$-phenylenevinylene) (PPV) are theoretically investigated within a coarse-grained model, in which both the exciton and nuclear degrees of freedom are treated quantum mechanically. The Frenkel-Holstein Hamiltonian is used to describe the strong exciton-phonon coupling present in the system, while external damping of the internal nuclear degrees of freedom are accounted for by a Lindblad master equation. Numerically, the dynamics are computed using the time evolving block decimation (TEBD) and quantum jump trajectory techniques. The values of the model parameters physically relevant to polymer systems naturally lead to a separation of time scales, with the ultra-fast dynamics corresponding to energy transfer from the exciton to the internal phonon modes (i.e., the C-C bond oscillations), while the longer time dynamics correspond to damping of these phonon modes by the external dissipation. Associated with these time scales, we investigate the following processes that are indicative of the system relaxing onto the emissive chromophores of the polymer: 1) Exciton-polaron formation occurs on an ultra-fast time scale, with the associated exciton-phonon correlations present within half a vibrational time period of the C-C bond oscillations. 2) Exciton decoherence is driven by the decay in the vibrational overlaps associated with exciton-polaron formation, occurring on the same time scale. 3) Exciton density localization is driven by the external dissipation, arising from `wavefunction collapse' occurring as a result of the system-environment interactions. Finally, we show how fluorescence anisotropy measurements can be used to investigate the exciton decoherence process during the relaxation dynamics.
\end{abstract}

\maketitle

\section{Introduction}
Upon photoexcitation of a conjugated polymer with a pulse of electromagnetic radiation, the exciton state formed is no longer stabilised by the ground state nuclear geometry, leading to coupled exciton-phonon dynamics as the system relaxes back to equilibrium. These dynamics have been investigated using a wide array of time-resolved spectroscopic techniques for the polymer poly($p$-phenylenevinylene) (PPV), including fluorescence depolarization,\cite{fluor,fluor1,fluor2} three-pulse photon-echo \cite{echo,echo1,echo2} and coherent electronic two-dimensional spectroscopy,\cite{2d} with multiple time scales being identified. Based on experimental and theoretical studies,\cite{diffusion,diffusion1,fluor1} the largest time scale seen in these experiments of ${>}1$~ps is widely accepted to correspond to F\"orster type exciton energy transfer between chromophores. The two shorter time scales of ${\lesssim}50$~fs and ${\sim}100$~fs, however, have been studied in less detail. While it has been suggested that these two time scales correspond to the dynamic localization of the exciton as it relaxes to the low energy chromophores,\cite{fluor,fluor1,fluor2,echo,echo1,echo2,2d} to our knowledge this has yet to be confirmed by theoretical simulations.

One reason for this lack of theoretical insight is the inherent challenge in simulating the exciton relaxation dynamics numerically, due to the strong exciton-phonon interactions synonymous with polymer systems. One common way to treat exciton-phonon interactions within these systems is to use the one dimensional Frenkel-Holstein model,\cite{holst1, *holst2,frenk_holst} which consists of a linear array of sites through which the exciton can propagate (where each site corresponds to a moiety in the polymer chain), and where the nuclear degrees of freedom are represented by a single harmonic oscillator per site, which couples locally to the exciton. The size of the Hilbert space associated with this model grows exponentially with the number of sites in the chain, which means that computationally solving the time-dependent Schr\"odinger equation using standard differential equation solvers becomes unfeasible for even modest chain lengths.

One way to deal with the exponentially increasing Hilbert space is to treat the nuclear degrees of freedom classically, within the so-called Ehrenfest approximation. The Hilbert space now only contains the exciton degrees of freedom and therefore grows linearly with the number of moieties in the polymer chain, making it computationally feasible to perform exciton dynamics calculations for experimentally relevant polymer chain lengths. Indeed, this technique has been applied to study the exciton relaxation dynamics of photoexcited electronic states in PPV.\cite{ehren1,ehren2,ehren3} However, the Ehrenfest approximation fails because of the absence of exciton decoherence occurring during the time evolution induced by the exciton-phonon coupling,\cite{decohere,overlap,overlap1,overlap2,branch} as well as its inability to correctly describe the splitting of a nuclear wave packet when passing through a conical intersection or avoided crossing.\cite{branch,branch1} In fact, the latter is the cause of the unphysical bifurcation of the exciton density onto separate chromophores, found to occur during the relaxation dynamics of high energy photoexcited states modeled previously within the Ehrenfest approximation.\cite{ehren1} A correct description of the coupled exciton-phonon dynamics therefore requires a full quantum mechanical treatment of the system, as described in this paper.

For one dimensional strongly correlated systems, the density matrix renormalization group technique has been successful in obtaining accurate numerical results for properties associated with the ground state of the system.\cite{dmrg,mps} Using the same Hilbert space truncation scheme, the time evolving block decimation (TEBD) technique\cite{tebd,tebd1,tebd2} is an efficient and accurate method for computing the short time dynamics of such systems. One appealing feature of this method is that the associated computational cost scales linearly with the number of `sites' in the system,\cite{tebd,tebd2} allowing in principle the quantum dynamics for large systems to be simulated. Indeed, the technique has already been applied to obtain accurate results for the charge dynamics within the one dimensional Holstein model.\cite{dynam_holst,dynam_holst1}

Exciton relaxation dynamics for photoexcited states of PPV are modeled in this paper using the Frenkel-Holstein model and TEBD method. The phonon modes to which the exciton is coupled are assumed to be those associated with the high frequency C-C bond oscillations. The low frequency torsional modes are neglected (although their possible role in relaxation and decoherence is discussed in Sec.~\ref{sect:conclusion}). We model environmental and conformational static disorder through the exciton on-site energies and hopping integrals, which leads to the exciton density associated with the low energy eigenstates of this model being spatially Anderson localized,\cite{anderson} thus defining the emissive chromophores of the polymer. 
External damping of the internal phonon modes is included within a Lindblad master equation approach, which allows the system to dissipate energy to the environment and thus relax into the low energy eigenstates.

This paper is structured as follows. In Sec.~\ref{sect:conformation}, we show how a polymer chain can be represented by a coarse-grained model, generating a reduced dimensional Frenkel exciton Hilbert space in which to model the associated dynamics. In Sec.~\ref{sect:models}, the Frenkel-Holstein Hamiltonian is introduced for this coarse-grained model, which describes the important interactions associated with the exciton and nuclear degrees of freedom. In addition, the Lindblad master equation is introduced along with the numerical techniques used to solve the underlying equations. We then present our results in Sec.~\ref{sect:results}, including a discussion of the various processes associated with the short time relaxation dynamics, such as exciton-polaron formation, exciton decoherence and exciton density localization. We also make a connection between these processes and fluorescence depolarization measurements. Finally, we conclude in Sec.~\ref{sect:conclusion}, with a particular emphasis on comparing our results to experiment.

\section{Polymer Conformations and Coarse-Graining}
\label{sect:conformation}
Polymers exist in many conformations, arising from fluctuations in the torsional angles around the single carbon-carbon bonds, as well as defects in the polymer chains themselves. While in the solid state these torsional angle fluctuations are quasi-static, in solution they give rise to dynamical disorder. However, as the time scale for rotation around these single carbon-carbon bonds is much longer than the associated time scale for exciton relaxation, the observables associated with the exciton relaxation dynamics can be calculated by averaging over many different static polymer chain conformations for both phases.

\begin{figure}
\includegraphics[width=7cm]{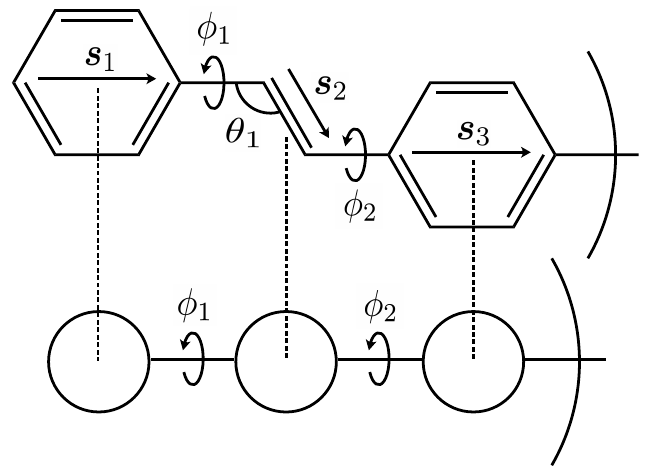}
\caption{\label{fig:conformation} The mapping of a polymer chain conformation to a coarse-grained linear site model. Each site corresponds to a moiety along the polymer chain, with the connection between sites characterised by the torsional angle, $\phi_{n}$. Also shown is the dipole unit vector, $\boldsymbol{s}_{n}$, associated with site/moiety $n$.}
\end{figure}

Figure~\ref{fig:conformation} illustrates how each polymer conformation of PPV is generated. The polymer chain is built up from the starting moiety, a phenylene unit, using the parameter values given in Table~\ref{tab:param_conf}, along with the following rules:
\begin{itemize}
\item The torsional angles, $\phi_{n}$, are taken as Gaussian random variables, with a mean $\langle\phi\rangle$ and standard deviation $\sigma_{\phi}$. 
\item With equal probability, the torsional angles, $\phi_{n}$, and the vinylene bond angles, $\theta_{n}$, can be positive or negative, corresponding to an anticlockwise or clockwise rotation respectively.
\item For a pure polymer chain, all vinylene units take the lowest energy trans geometry. However, trans-cis defects are introduced randomly along the chain, with probability $x_{\text{cis}}$.
\end{itemize}

In general, polymers contain many thousands of atoms and therefore performing a full atomistic simulation of exciton dynamics for each polymer conformation is unfeasible. Our model can be coarse-grained, as has been done in previous work,\cite{ehren1} by representing a polymer chain as a linear array of sites, where each site corresponds to one of the moieties of the polymer. This mapping is illustrated in Fig.~\ref{fig:conformation}. 

In conjugated polymer systems, such as PPV, the large electron-electron interactions lead to the electron-hole pair of the exciton being tightly bound, typically occupying the same moiety of the chain.\cite{frenkel} This is known as a Frenkel exciton. Thus in our coarse-grained model, we only need to consider how the center of mass of the Frenkel exciton propagates along the array of sites. A single exciton state is kept per site, corresponding to the lowest energy excitation of each moiety.
\begin{table}[H]
\centering
{\renewcommand{\arraystretch}{1.2}
\begin{tabular}{|M{2cm}|M{2cm}|M{2cm}|M{2cm}|}
\hline
$\textbf{Parameter}$ & $\textbf{Value}$ & $\textbf{Parameter}$ & $\textbf{Value}$ \\ \hline
$\langle\phi\rangle$ & $15\degree$ & $\sigma_{\phi}$ & $5\degree$ \\
$x_{\text{cis}}$ & 0.08 & $\theta$ & $60\degree$ \\
\hline
\end{tabular}}
\caption{Parameters used to generate the conformations associated with PPV polymer chains.}
\label{tab:param_conf}
\end{table}	

\section{Models and Numerical Techniques}
\label{sect:models}
In this section we introduce the model Hamiltonians, as well as outlining the various numerical techniques used to solve the associated exciton dynamics.

\subsection{The Frenkel Model}
In previous work, the initial exciton state of a PPV polymer, generated after photoexcitation, has been described by the Frenkel Hamiltonian:\cite{book}
\begin{equation}
\label{eq:frenkel}
\hat{H}_{\text{F}} = \sum_{n}\epsilon_{n}\hat{a}_{n}^{\dagger}\hat{a}_{n} + \sum_{n}J_{n}\left(\hat{a}_{n+1}^{\dagger}\hat{a}_{n}+\hat{a}_{n}^{\dagger}\hat{a}_{n+1}\right)
\end{equation}
where $\hat{a}_{n}^{\dagger}$ ($\hat{a}_{n}$) is the exciton creation (destruction) operator, which creates (destroys) a Frenkel exciton on moiety~$n$ of the polymer chain. For PPV, the odd and even sites, $n$, correspond to phenylene and vinylene moieties respectively. Within this Hamiltonian, the on-site energy is given by:
\begin{equation}
\label{eq:on-site}
\epsilon_{n} = E_{0} + \left(-1\right)^{n}\frac{\Delta}{2} + \alpha_{n}
\end{equation}
where $E_{0}$ is the average moiety excitation energy, $\Delta$ is the difference in the excitation energy between phenylene and vinylene moieties and $\alpha_{n}$ represents the diagonal disorder, with standard deviation $\sigma_{\alpha}$. This diagonal disorder arises physically from density fluctuations in the environment around the polymer chain, due to the inhomogeneity of the material.

Additionally, the nearest neighbor exciton hopping integrals present in the Frenkel Hamiltonian are given by:\cite{dis_off,dis_off2}
\begin{equation}
\label{eq:hopping}
J_{n} = J^{\text{DD}} + J^{\text{SE}}\cos^{2}{\phi_{n}}
\end{equation}
where $\phi_{n}$ are the torsional angles between moieties along the polymer chain. From Eq.~(\ref{eq:hopping}), we see that there are two contributions to the nearest neighbor hopping integral. The first contribution arises from a through space dipole-dipole interaction, which is incorporated through the term $J^{\text{DD}}$. While in principle this interaction term is non zero between all moieties in the polymer chain, for this analysis we only keep the nearest neighbor interactions, which are the dominant terms. The second contribution to the hopping integral arises from a through bond super exchange interaction, in which the system passes through an intermediate charge transfer exciton state. The strength of this interaction depends on the torsional angle between the moieties ($\propto\cos^{2}{\phi_{n}}$, where $\cos{\phi_{n}}$ gives the overlap of the $p_{z}$ orbitals on each moiety). Thus torsional fluctuations in the polymer chain give rise to disordered hopping integrals in the Frenkel Hamiltonian.

\begin{figure}
\includegraphics[width=8.5cm]{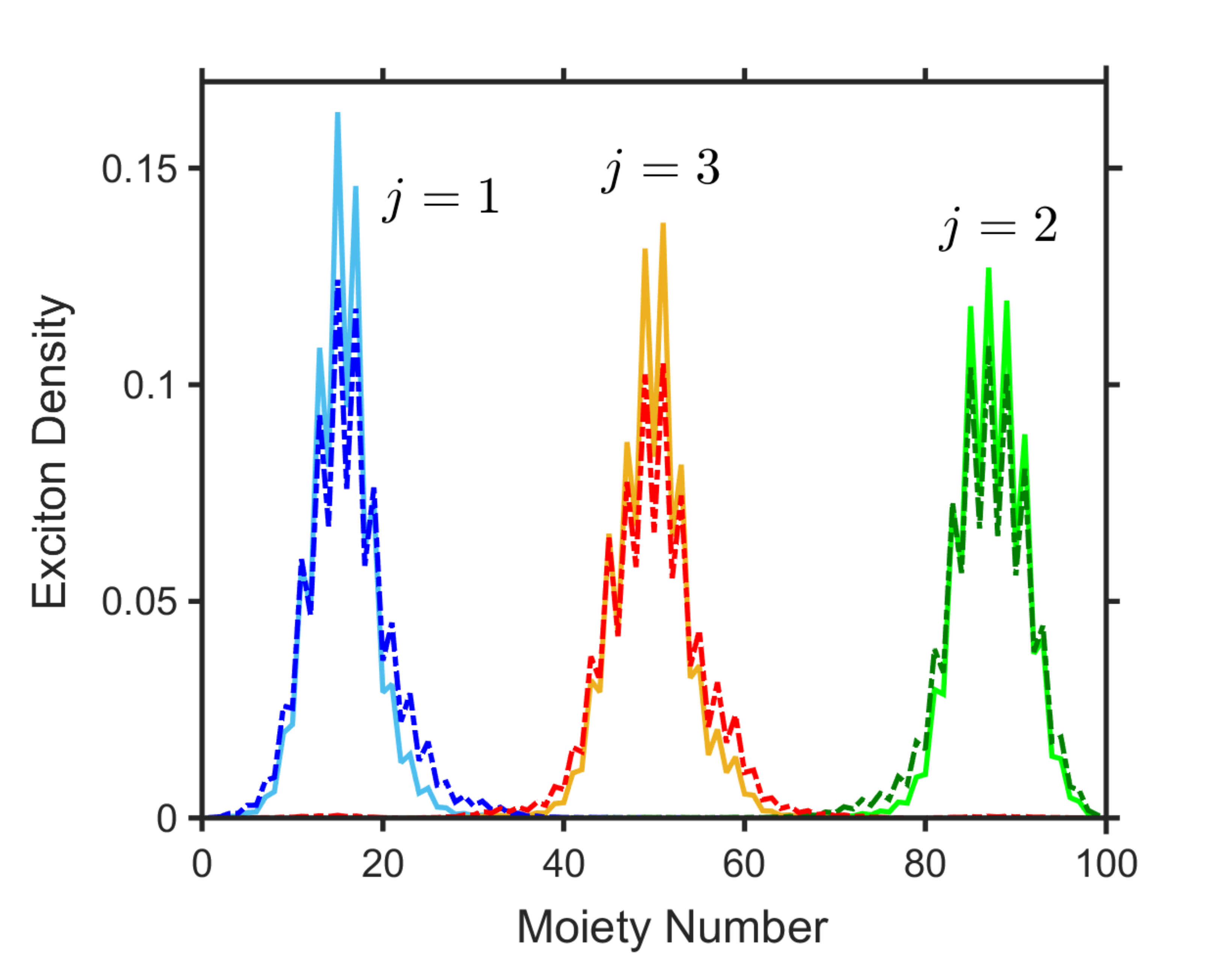}
\caption{\label{fig:dmrg} The three LEGSs (dotted lines) and three VRSs (solid lines) for one particular conformation of a PPV polymer chain made up of 99 moieties. The exciton center-of-mass quantum number, $j$, for each state is also given. The VRSs were obtained using the density matrix renormalization group technique, allowing a maximum of two phonons per site.}
\end{figure}

\begin{figure}
\includegraphics[width=8.5cm]{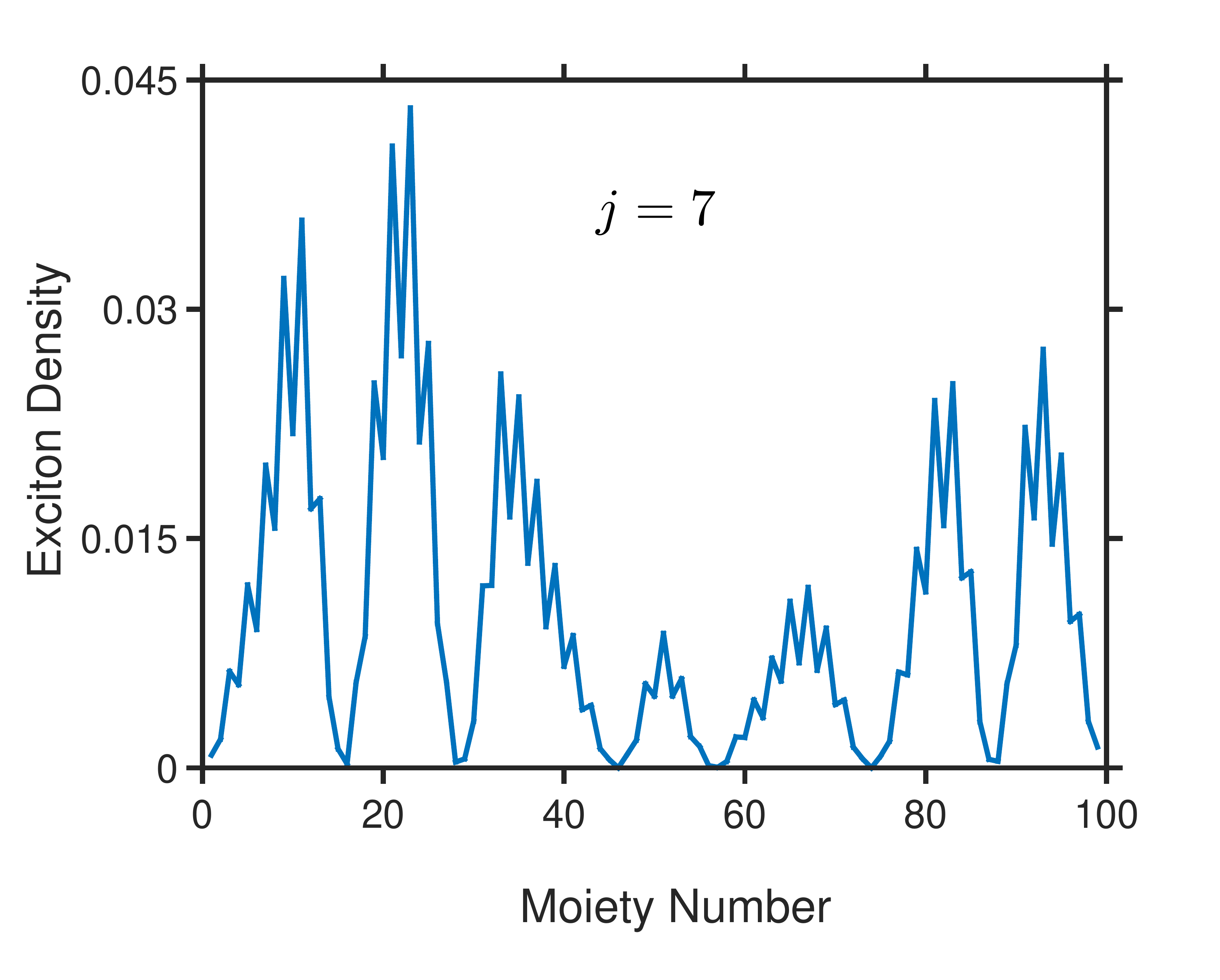}
\caption{\label{fig:qees} The associated exciton density of a QEES, with quantum number $j=7$, for one particular conformation of a PPV polymer chain made up of 99 moieties.}
\end{figure}

The disorder present in the model leads to the Frenkel Hamiltonian having two forms of eigenstates. The low energy states of the Hamiltonian are Anderson localized, non-overlapping and nodeless states, called local exciton ground states (LEGSs).\cite{legs1,legs2} A LEGS is quantified by a signed-value parameter, $\alpha$, defined as:
\begin{equation}
\alpha=\left|\sum_{n}\left|\psi_{n}\right|\psi_{n}\right|
\end{equation}
where $\psi_{n}$ is the probability amplitude of the Frenkel exciton being on site $n$. In accord with previous work, we define a LEGS as satisfying $\alpha\geq0.95$.\cite{legs1,legs2} The exciton density for three LEGSs corresponding to a particular conformation of a 99 moiety PPV polymer chain are given by the dotted curves in Fig.~\ref{fig:dmrg}. It is the width of these states that define the size of the polymer's absorbing chromophores.\cite{chromophore} In addition, the Frenkel Hamiltonian has higher energy eigenstates that are delocalized over several chromophores, called quasi-extended exciton states (QEES).\cite{book} The exciton density for a QEES on the same 99 moiety PPV polymer chain is given in Fig.~\ref{fig:qees}.

\subsection{The Frenkel-Holstein Model}
Relaxation of a LEGS or QEES is physically induced by exciton-nuclear coupling, which allows the nuclear geometry of the polymer to distort so that it best stabilises the newly formed excited electronic state. Most vibrational normal modes in a polymer such as PPV either couple weakly to the exciton or have a low vibrational frequency, such that they do not strongly influence the exciton dynamics on the ultra-fast time scale of interest and therefore can be neglected in our model. As in previous work, we thus consider the Frenkel exciton coupling to a single high frequency normal mode per moiety (the benzenoid-quinoid distortion in the phenylene unit and the C-C double bond stretch in the vinylene unit) modeled using the Frenkel-Holstein Hamiltonian:\cite{holst1, *holst2,frenk_holst}
\begin{equation}
\label{eq:holstein}
\hat{H}_{\text{FH}} = \hat{H}_{\text{F}} - \frac{A\hbar\omega}{\sqrt{2}}\sum_{n}\hat{a}_{n}^{\dagger}\hat{a}_{n}\left(\hat{b}_{n}^{\dagger}+\hat{b}_{n}\right)+\hbar\omega\sum_{n}\left(\hat{b}_{n}^{\dagger}\hat{b}_{n}+\tfrac{1}{2}\right)
\end{equation}
where $A$ is the dimensionless exciton-nuclear coupling parameter, which determines the strength of the exciton-phonon coupling, and $\hbar\omega$ is the phonon energy associated with the vibrational normal modes. Within the Frenkel-Holstein model, the vibrational normal mode on moiety/site $n$ is represented by a harmonic oscillator, where $\hat{b}_{n}^{\dagger}$ ($\hat{b}_{n}$) is the associated dimensionless harmonic oscillator creation (destruction) operator.

One effect of the exciton-nuclear coupling term in this Hamiltonian is to further localize the exciton density of the low energy LEGSs, to create so-called vibrationally relaxed states (VRSs).\cite{polaron_landau} These VRSs are the low energy eigenstates of the Frenkel-Holstein Hamiltonian,\cite{polaron_holst} with their associated exciton densities given by the solid curves in Fig.~\ref{fig:dmrg} for a 99 moiety PPV polymer chain. Comparing these VRSs with their associated LEGSs (given by the dotted curves) shows that the degree of self-localization arising from coupling to the high frequency vibrational modes is small, leading to the spatial extent of the VRSs still being largely determined by the disorder in the Frenkel Hamiltonian.\cite{polaron_holst,polaron_landau} It is the width of these VRSs that define the size of the polymer's emissive chromophores.\cite{ehren1}

The values for the Frenkel-Holstein Hamiltonian parameters that we used to model PPV\cite{ehren1} are given in Table~\ref{tab:parameters}.
\begin{table}[H]
\centering
{\renewcommand{\arraystretch}{1.3}
 \begin{tabular}{|M{2cm}|M{2cm}|M{2cm}|M{2cm}|}
\hline
$\textbf{Parameter}$ & $\textbf{Value}$ & $\textbf{Parameter}$ & $\textbf{Value}$ \\ \hline
$E_{0}$ & $9.24$~eV & $J^{\text{DD}}$ & $-1.35$~eV \\
$\Delta$ & $3.20$~eV & $\hbar\omega$ & $0.2$~eV \\
$\sigma_{\alpha}$ & $65$~meV & $A$ & $4.00$ \\
$J^{\text{SE}}$ & $-1.96$~eV & $\tilde{\gamma}$ & 0.033 \\
\hline
\end{tabular}}
\caption{Parameters used in the Frenkel-Holstein Hamiltonian.}
\label{tab:parameters}
\end{table}	

\subsection{The Lindblad Master Equation}\label{Se:III.C}
To describe the time evolution of a Frenkel exciton (initially in a high energy QEES or LEGS) relaxing to the low energy VRSs, the model must allow the system to lose excess energy to the environment. Dissipation of energy arising from system-environment coupling is commonly accounted for by describing the time evolution of the quantum system by a Lindblad master equation:\cite{open}
\begin{equation}
\label{eq:master}
\frac{\partial\hat{\rho}}{\partial \tilde{t}} = -\frac{i}{\hbar\omega}\left[\hat{H},\hat{\rho}\right]-\frac{\tilde{\gamma}}{2}\sum_{n}\left(\hat{b}_{n}^{\dagger}\hat{b}_{n}\hat{\rho}+\hat{\rho}\hat{b}_{n}^{\dagger}\hat{b}_{n}-2\hat{b}_{n}\hat{\rho}\hat{b}_{n}^{\dagger}\right)
\end{equation}
where $\tilde{t}=t\omega$ is the dimensionless time. Such an equation describes the time dependence of the system density matrix, $\hat{\rho}$, from which any observable associated with the system of interest can be obtained. From Eq.~(\ref{eq:master}), we see that the time dependence of the system density matrix depends on two terms. The first term on the right of Eq.~(\ref{eq:master}) corresponds to the evolution of the system density matrix under the Hamiltonian, $\hat{H}$, given by:
\begin{equation}
\label{eq:ham_correc}
\hat{H}=\hat{H}_{\text{FH}}+\frac{\tilde{\gamma}\hbar\omega}{4}\sum_{n}\left(\hat{Q}_{n}\hat{P}_{n}+\hat{P}_{n}\hat{Q}_{n}\right)
\end{equation}
where $\hat{Q}_{n}=\tfrac{1}{\sqrt{2}}(\hat{b}_{n}^{\dagger}+\hat{b}_{n})$ is the dimensionless displacement operator and $\hat{P}_{n}=\tfrac{i}{\sqrt{2}}(\hat{b}_{n}^{\dagger}-\hat{b}_{n})$ is the dimensionless momentum operator, for the oscillator on site $n$, and the last term in Eq.~(\ref{eq:ham_correc}) accounts for the damping correction to the harmonic oscillator frequency induced by the external dissipation.\cite{brownian} The second term in Eq.~(\ref{eq:master}) accounts for energy dissipation of the system induced by the system-environment coupling. The effect of the system-environment coupling is controlled by the dimensionless dissipation parameter $\tilde{\gamma}=\frac{\gamma}{\omega}$, as well as the Lindblad operator for site $n$, $\hat{b}_{n}$, which for the system considered in this paper becomes the dimensionless harmonic oscillator destruction operator.

While the Lindblad master equation only approximately describes dissipation effects due to system-environment coupling, the perturbative nature of its derivation means that we expect the master equation to describe the dynamics of systems in which the system-environment coupling is weak (i.e., when $\tilde{\gamma}$ is small). Such a situation is indeed physically relevant to the exciton relaxation dynamics in polymer chains, in which the exciton-phonon coupling strength is much larger that any coupling terms to the environment.

It is not necessarily obvious why the harmonic oscillator destruction operator is a good choice for the Lindblad operator in our master equation in Eq.~(\ref{eq:master}). To show that it is, we first note that the expectation value of some operator corresponding to a system observable, $\hat{O}$, can be calculated from the system density matrix as follows:
\begin{equation}
\label{eq:expec}
\langle\hat{O}\rangle=\text{Tr}\left[\hat{\rho}\hat{O}\right]
\end{equation}
where $\text{Tr}[\dotsm]$ corresponds to taking the trace of the quantity inside the brackets. Using Eq.~(\ref{eq:master}), we can then derive the following equations of motion for the dimensionless nuclear displacement and momentum:
\begin{align}
\label{eq:Classical}
\begin{split}
\frac{d\langle\hat{Q}_{n}\rangle}{d\tilde{t}} & = \langle\hat{P}_{n}\rangle \\
\frac{d\langle\hat{P}_{n}\rangle}{d\tilde{t}} & = A\langle\hat{a}_{n}^{\dagger}\hat{a}_{n}\rangle - \langle\hat{Q}_{n}\rangle - \tilde{\gamma}\langle\hat{P}_{n}\rangle
\end{split}
\end{align}
which have the form of Newton's equations of motion for a damped harmonic oscillator \cite{open,brownian} and are the identical equations of motion used previously for modeling exciton relaxation dynamics for classical nuclei.\cite{ehren1}

\subsection{Optical Intensity}
In an experiment, the initial exciton state from which the relaxation dynamics proceed is formed by photoexcitation (typically with a broadband pulse) from the electronic ground state. On application of this pulse, the nuclear degrees of freedom remain static (i.e., the harmonic oscillators remain in their ground state), while the exciton is created in a state that corresponds to a linear combination of the eigenstates of the Frenkel Hamiltonian. The probability that an eigenstate of the Frenkel Hamiltonian with energy $E$ contributes to this initial exciton state is given by the spectral function:
\begin{equation}
\label{eq:optical_spectrum}
I\left(E\right)=\left\langle\sum_{\alpha}f_{\alpha}\delta\left(E-E_{\alpha}\right)\right\rangle_{\text{Disorder}}
\end{equation}
where $E_{\alpha}$ is the energy eigenvalue associated with eigenstate $|\psi_{\alpha}\rangle$ of the Frenkel Hamiltonian. In addition, $f_{\alpha}$ is the associated oscillator strength for this eigenstate, given by:
\begin{equation}
\label{eq:oscillator_strength}
f_{\alpha} = \left(\frac{2m_{e}E_{\alpha}}{3e^{2}\hbar^{2}}\right)
\sum_{s=x,y,z}\left|\braket{\psi_{\alpha}|\hat{\mu}_{s}|0}\right|^{2}
\end{equation}
where $\ket{0}$ corresponds to the electronic ground state of the system. On calculating the spectral function using Eq.~(\ref{eq:optical_spectrum}), we average our result over the various instances of the disorder in the Frenkel Hamiltonian, as well as over all possible conformations of the polymer chain. For a particular conformation, the transition dipole moment operator, $\boldsymbol{\hat{\mu}}$, takes the form:
\begin{equation}
\label{eq:dipole}
\boldsymbol{\hat{\mu}}= \mu_{0}\sum_{n}\boldsymbol{s}_{n}\left(\hat{a}_{n}^{\dagger}+\hat{a}_{n}\right)
\end{equation}
where $\mu_{0}$ is the magnitude of the transition dipole moment for a Frenkel exciton localized on a single moiety of the polymer and $\hat{a}_{n}^{\dagger}$ ($\hat{a}_{n}$) is the exciton creation (destruction) operator for site/moiety~$n$. Additionally, $\boldsymbol{s}_{n}$ is a unit vector that points along the polymer axis at moiety~$n$, which is illustrated in Fig.~\ref{fig:conformation}.

\begin{figure}
\includegraphics[width=8.5cm]{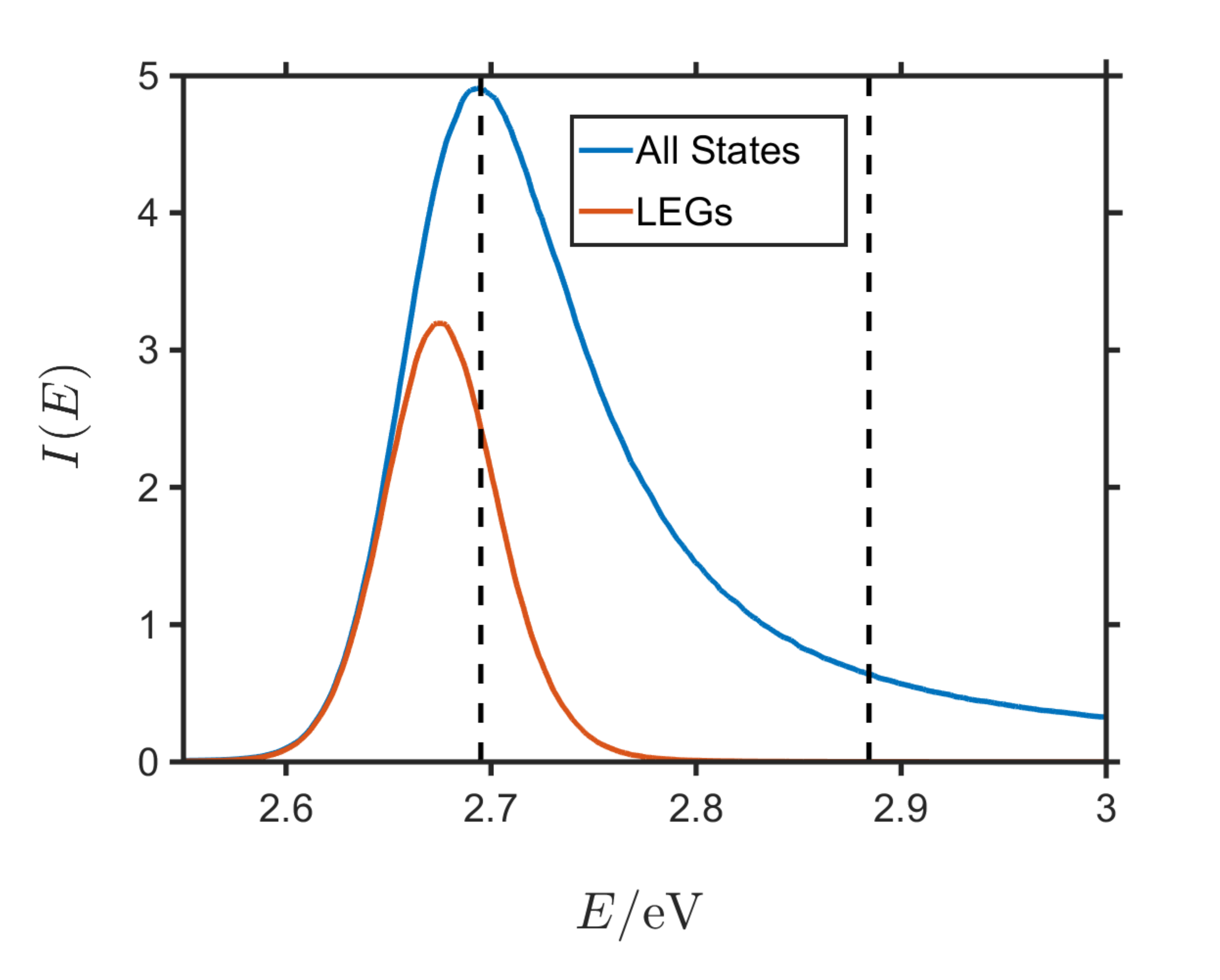}
\caption{\label{fig:optical} The theoretically obtained spectral function, $I(E)$, for PPV polymers, that corresponds to the probability that Frenkel exciton states with energy $E$ are produced by a broadband photoexcitation pulse. The low energy part of the spectrum is dominated by the LEGSs, whose spatial extent define the size of the polymers absorbing chromophores. The dotted lines at $2.68$~eV and $2.88$~eV illustrate the initial Frenkel exciton energies from which the exciton relaxation dynamics are simulated.}
\end{figure}

Figure~\ref{fig:optical} shows the calculated spectral function for PPV, using the parameter values given in Tables~\ref{tab:param_conf}~and~\ref{tab:parameters}. In this work, we chose to perform our relaxation dynamics for initial Frenkel exciton energies of 2.68~eV and 2.88~eV, shown by the dotted lines in Fig.~\ref{fig:optical}. Both of these Frenkel exciton energies have appreciable formation probabilities and also allow us to investigate the differing relaxation dynamics of LEGSs, low energy QEESs and high energy QEESs.

\subsection{Numerical Techniques}
\label{sect:numerical}
Section~\ref{Se:III.C} outlines the necessary theoretical framework for modeling the exciton relaxation dynamics. However, evolving the system density matrix numerically using the Lindblad master equation given in Eq.~(\ref{eq:master}) is not straight forward for the following reasons. The Frenkel-Holstein model contains exciton-phonon interactions, which lead to its associated Hilbert space growing exponentially with the number of sites/moieties in the polymer chain. Thus for chains containing a physically realistic number of moieties/sites, this Hilbert space is too large to manipulate on a computer. This issue is compounded further, as the size of the system density matrix is the Hilbert space squared, meaning the use of standard differential equation solvers to compute the system density matrix from Eq.~(\ref{eq:master}) is intractable. In this work, these issues are overcome by using the time evolving block decimation (TEBD)\cite{tnt,tebd1,tebd2} and the quantum jump trajectory methods,\cite{jump,jump1,jump2,jump3} respectively. The details of these techniques are not discussed here, but since these are unfamiliar to the chemical physics community, an overview of both can be found in the Appendices.

Before a discussion of our results, we briefly outline the basic constituents of the quantum jump trajectory method, which will prove useful in interpreting the role of the external dissipation in our calculated dynamics. The quantum jump trajectory method obtains system observables associated with the numerical solution of a Lindblad master equation, such as Eq.~(\ref{eq:master}), by averaging over so-called quantum trajectories. A single trajectory is represented by a state in the system Hilbert space, with its time evolution determined as follows. Typically, the trajectory evolves under the effective Hamiltonian:\cite{jump}
\begin{equation}
\label{eq:heff}
\hat{H}_{\text{eff}}=\hat{H}-\frac{i\tilde{\gamma}}{2}\sum_{n}\hat{b}_{n}^{\dagger}\hat{b}_{n}
\end{equation}
where $\hat{H}$ is the system Hamiltonian given in Eq.~(\ref{eq:ham_correc}). However, at probabilistically determined times, the trajectory undergoes a `quantum jump', which corresponds to application of one of the Lindblad operators onto the state. Averaging over a large number of these stochastically determined trajectories is known to accurately reproduce the time-dependent observables associated with the Lindblad master equation.\cite{jump}

\section{Results}
\label{sect:results}
\begin{figure}
\includegraphics[width=8.5cm]{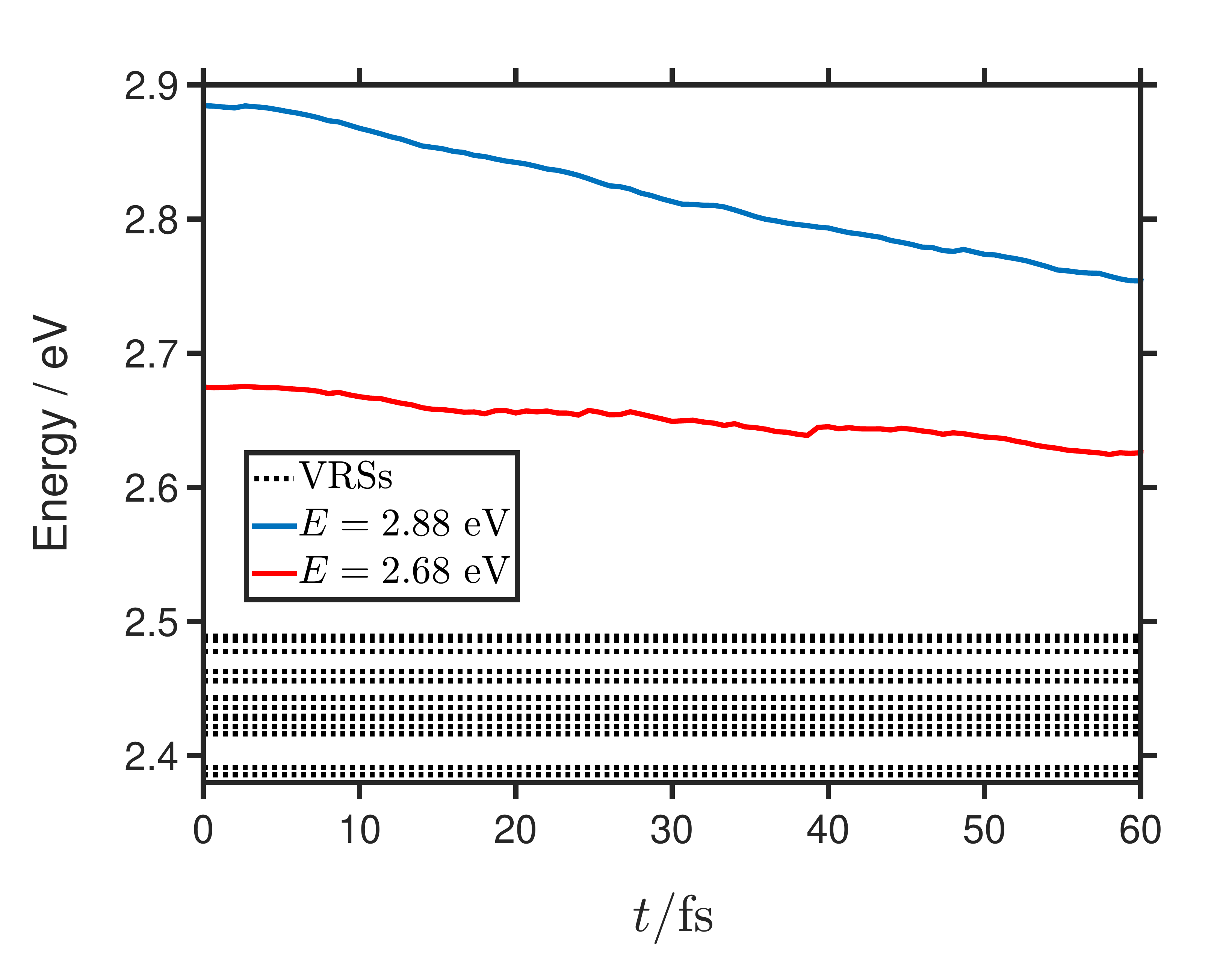}
\caption{\label{fig:energy_tot} The time dependence of the total system energy. The blue curve corresponds to an initial Frenkel exciton energy of $2.88$~eV, while the red curve corresponds to an initial Frenkel exciton energy of $2.68$~eV. The black dashed lines show the corresponding system energy for the VRSs of the polymer chain for different instances of the disorder.}
\end{figure}

The dynamics must fulfil certain criteria if the time evolution determined by our theoretical methodology is physically correct. One such requirement is that the total energy of the system must decrease during the dynamics, approaching the energies of the VRSs, which define the emissive chromophores of the polymer. We indeed find that the system energy does decrease in our dynamics model, as shown by the time evolution of the total energy of the system for two different initial Frenkel exciton energies given in Figure~\ref{fig:energy_tot}. This confirms that the form of the external dissipation in our model is having the desired effect. Also shown in this figure are the energies associated with the VRSs, for various instances of the disorder, given by the dotted curves. The fact that the total system energy at $t=60$~fs is greater than the energy of these VRSs for both initial Frenkel exciton energies, as well as $dE/dt$ being non zero at this time, suggests that the dynamics are far from equilibrium. This is not surprising, since the value of the exponential dissipation time scale, $\gamma^{-1}\sim100$~fs, means that the system will only reach equilibrium at times much longer than computationally practical simulations of ${\sim}100$~fs.

\begin{figure}
\includegraphics[width=8.5cm]{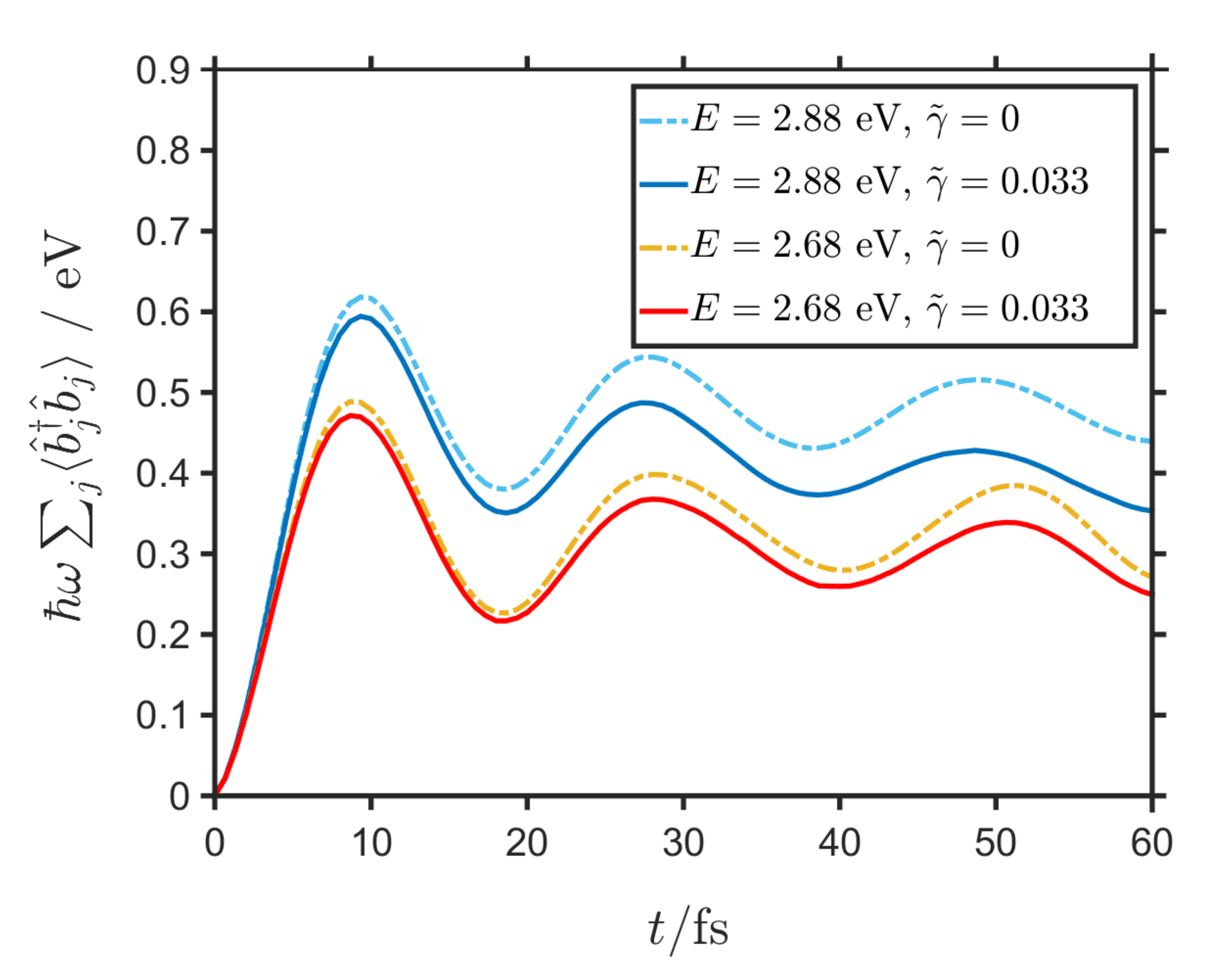}
\caption{\label{fig:energy_osc} The time dependence of the phonon energy. The blue curves correspond to an initial Frenkel exciton energy of $2.88$~eV, while the red curves correspond to an initial Frenkel exciton energy of $2.68$~eV. In addition, the dashed curves correspond to the relaxation dynamics with no energy dissipation to the environment, while the solid curves correspond to the dynamics with $\tilde{\gamma}=0.033$.}
\end{figure}

We observe a separation of time scales within our computed dynamics. This is illustrated in Fig.~\ref{fig:energy_osc}, which shows the time-dependent phonon energy for two different initial Frenkel exciton energies, with and without external dissipation. For very short times, the evolution of the system observables depend on the Frenkel-Holstein parameters and are largely independent of the external dissipation, $\tilde{\gamma}$. From Fig.~\ref{fig:energy_osc}, we see that the phonon energy initially increases during the dynamics, suggesting that this ultra-fast time scale is characterised by energy transfer from the exciton to the internal nuclear degrees of freedom (i.e., the phonon degrees of freedom act as a heat bath for the exciton). Eventually, the energy associated with the nuclear degrees of freedom saturates, with the time evolution of the observables now dependent on the longer external dissipation time scale. Figure~\ref{fig:energy_osc} shows that for $\tilde{\gamma}\neq0$, the phonon energy decreases at long times, suggesting that the long time scale dynamics are characterised by dissipation of the phonon energy to the environment (i.e., the environment acts as a heat bath for the internal nuclear degrees of freedom).

Even though the system is far from equilibrium for the times considered in our relaxation dynamics simulation, we should expect the time evolution of observables to still exhibit features associated with the steady states. Indeed, if the system does relax to the low energy VRSs as physically expected, processes such as exciton-polaron formation, exciton decoherence and exciton density localization should be present in the short time dynamics. We now consider each of these processes.

\subsection{Exciton-Polaron Formation}
\label{sect:polaron}
In systems with strong electron-phonon interactions, the low energy excitations of the system are described by polarons, which are quasiparticles consisting of an electron or exciton `dressed' or `self-trapped' by local displacements of the nuclei.\cite{polaron} (To distinguish dressed electrons and excitons, we use the term `exciton-polaron' in this paper for the dressed exciton.)
The effective size of the exciton-polaron can be described by the exciton-phonon correlation function:\cite{ex-ph-corr}
\begin{equation}
\label{eq:ex-ph-corr}
C_{n}^{\text{ex-ph}}=\frac{1}{A}\sum_{m}\braket{\hat{a}_{m}^{\dagger}\hat{a}_{m}\hat{Q}_{m+n}}
\end{equation}
where the factor $1/A$ normalizes the function for eigenstates of the Frenkel-Holstein Hamiltonian (i.e., \hbox{$\sum_{n}C_{n}^{\text{ex-ph}}=1$}).\cite{ex-ph-corr2} Physically, this correlation function corresponds to the average nuclear displacement $n$ sites away from the exciton. From previous work, it is known that for the VRSs of the Frenkel-Holstein model, the exciton-phonon correlation function has an exponential form, with a correlation length that is a function of the parameters $\hbar\omega$, $A$ and $\braket{J}$.\cite{polaron_holst} If the system does evolve to the VRSs, then we would expect the exciton-phonon correlation function to acquire this exponential behavior during the exciton relaxation dynamics.

\begin{figure}
\includegraphics[width=8.5cm]{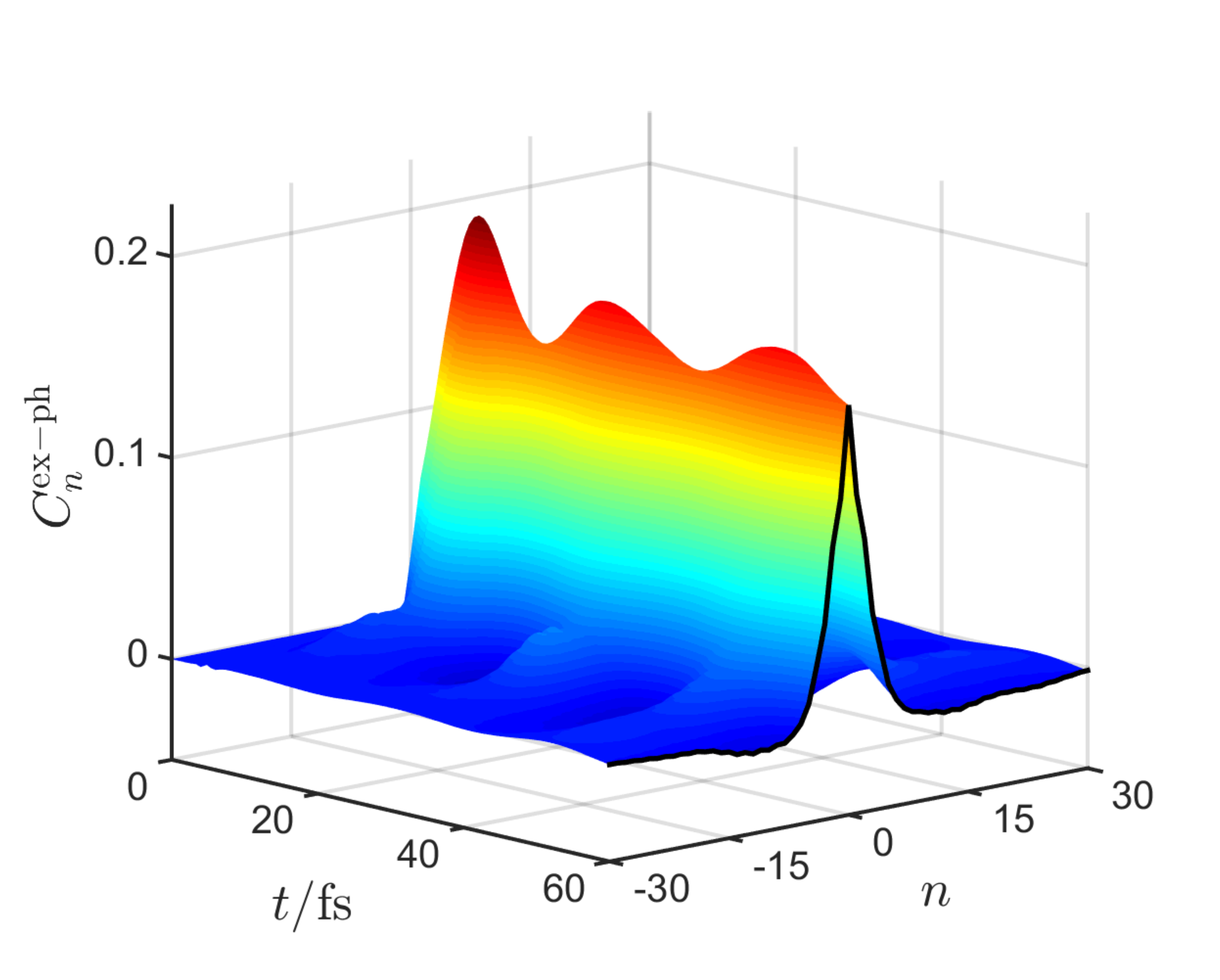}
\caption{\label{fig:corr_ex_ph} The time dependence of the exciton-phonon correlation function, $C_{n}^{\text{ex-ph}}$, for an initial high energy QEES given in Fig.~\ref{fig:qees}.}
\end{figure}
Figure~\ref{fig:corr_ex_ph} shows the time evolution of the exciton-phonon correlation function for an initial high energy QEES given in Fig.~\ref{fig:qees}. This correlation function acquires an exponential form during the dynamics, confirming that exciton-polaron formation is occurring within our model. Exciton-polaron formation occurs on an ultra-fast time scale (${\sim}10$~fs), with the dynamics being largely independent of the external dissipation. Indeed, we find that the time scale associated with exciton-polaron formation solely depends on the parameter $\hbar\omega$, with the local exciton-phonon correlations present within half a vibrational time period. In addition, persisting oscillations occur in the exciton-phonon correlation function with a time period (${\sim}20$~fs) that matches the vibrational time period of the harmonic oscillators in the model. These oscillations thus arise from the time evolution of the nuclear displacements $\hat{Q}_{n}$, which physically correspond to the C-C bond oscillations.

\begin{figure}
\includegraphics[width=8.5cm]{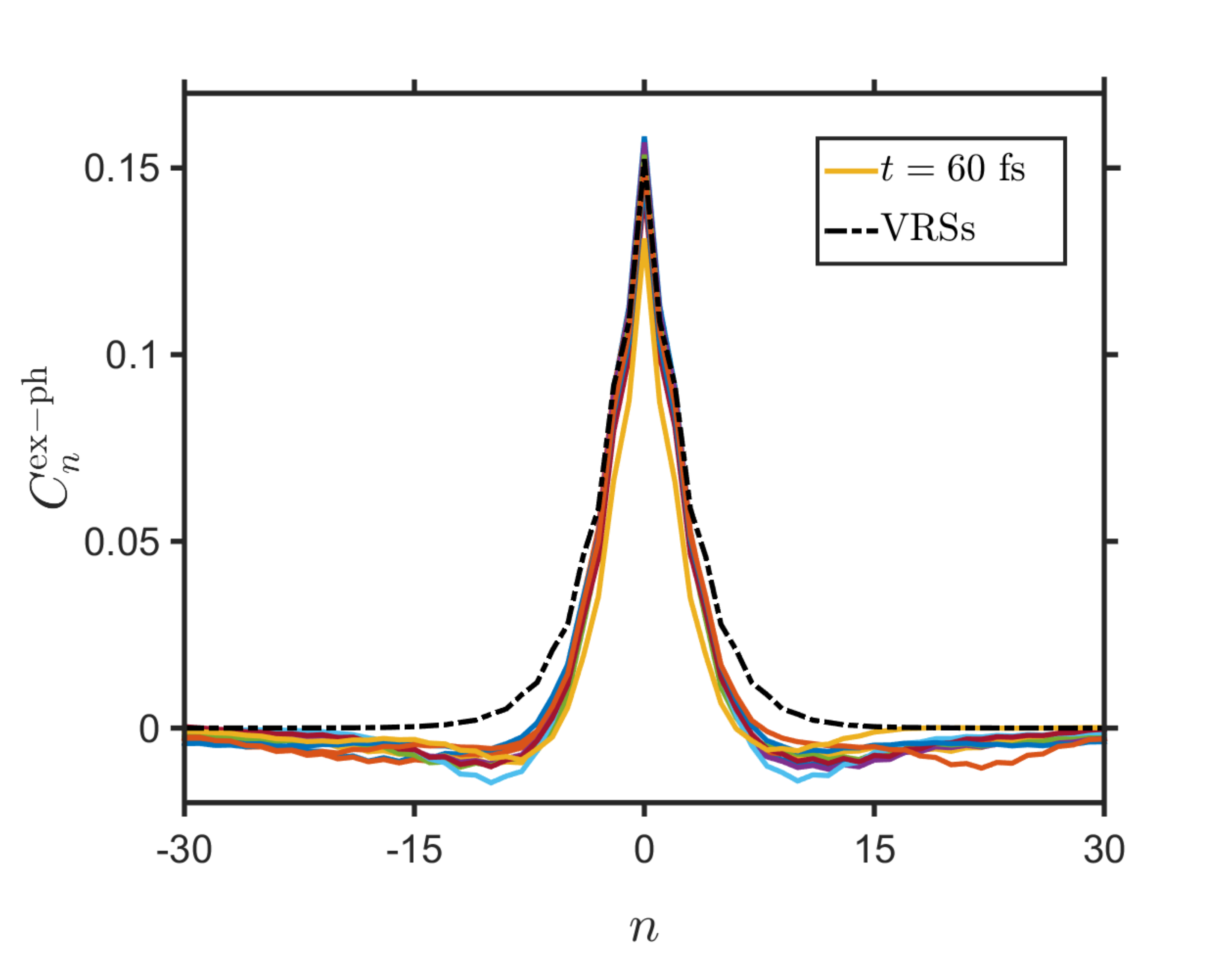}
\caption{\label{fig:corr_ex_ph2} The exciton-phonon correlation function, $C_{n}^{\text{ex-ph}}$, at $t=60$~fs for two initial photoexcitation energies ($2.68$~eV and $2.88$~eV) as well as several instances of the disorder in the Frenkel-Holstein Hamiltonian (given by the solid curves). Also shown is the average exciton-phonon correlation function for the VRSs, given by the dashed curve.}
\end{figure}
We now consider whether the correlation length associated with the dynamical exciton-phonon correlation function given in Fig.~\ref{fig:corr_ex_ph} is the same as for the VRSs. The solid curves in Fig.~\ref{fig:corr_ex_ph2}, which correspond to the exciton-phonon correlation functions at $t=60$~fs for two initial Frenkel exciton energies ($2.68$~eV and $2.88$~eV), as well as several instances of the disorder, all coincide. This is to be expected, as the parameters $\hbar\omega$, $A$ and $\braket{J}$ have the same values for all these curves, giving rise to an identical exciton-polaron formation time scale and exciton-phonon correlation length. Also shown in Fig.~\ref{fig:corr_ex_ph2} is the average exciton-phonon correlation function associated with the VRSs, given by the dashed curve. The form of the dynamical exciton-phonon correlation functions at $t=60$~fs shows good agreement with the same function for the VRSs, which is consistent with the system evolving to these states during the dynamics. However the agreement is not perfect, with deviations arising due to the persistent oscillations in the dynamical correlation function, which are symptomatic of the time evolution having not yet reached the steady state.

\subsection{Exciton Decoherence}
\label{sect:decohere}
An exciton confined to a one dimensional polymer chain in general can have long range quantum coherences between the moieties, which physically give rise to interference effects in the calculated observables. While the range of these coherences are often limited by the presence of disorder in the system, for QEESs the exciton coherences can still persist over a distance of several chromophores. If the exciton localizes onto a single chromophore during the relaxation dynamics, then these long range coherences will decay.

One way to quantify the magnitude of the exciton coherences is from the off-diagonal elements of the exciton reduced density matrix, $\hat{\rho}_{\text{ex}}$:
\begin{equation}
\label{eq:dens_ex}
\hat{\rho}_{\text{ex}}=\sum_{v}\braket{v|\hat{\rho}|v}
\end{equation}
which is obtained by taking the trace of the system density matrix over the nuclear degrees of freedom (characterised by the quantum number $v$). The following exciton coherence correlation function can then be defined, which gives the average magnitude of the exciton coherences between moieties/sites a distance $n$ apart:\cite{cohere1,cohere2}
\begin{equation}
\label{eq:ex-ex-corr}
C_{n}^{\text{coh}}=\sum_{m}\left|\braket{m|\hat{\rho}_{\text{ex}}|m+n}\right|
\end{equation}
where $\ket{m}$ is the ket corresponding to the exciton on site~$m$.

\begin{figure}
\includegraphics[width=8.5cm]{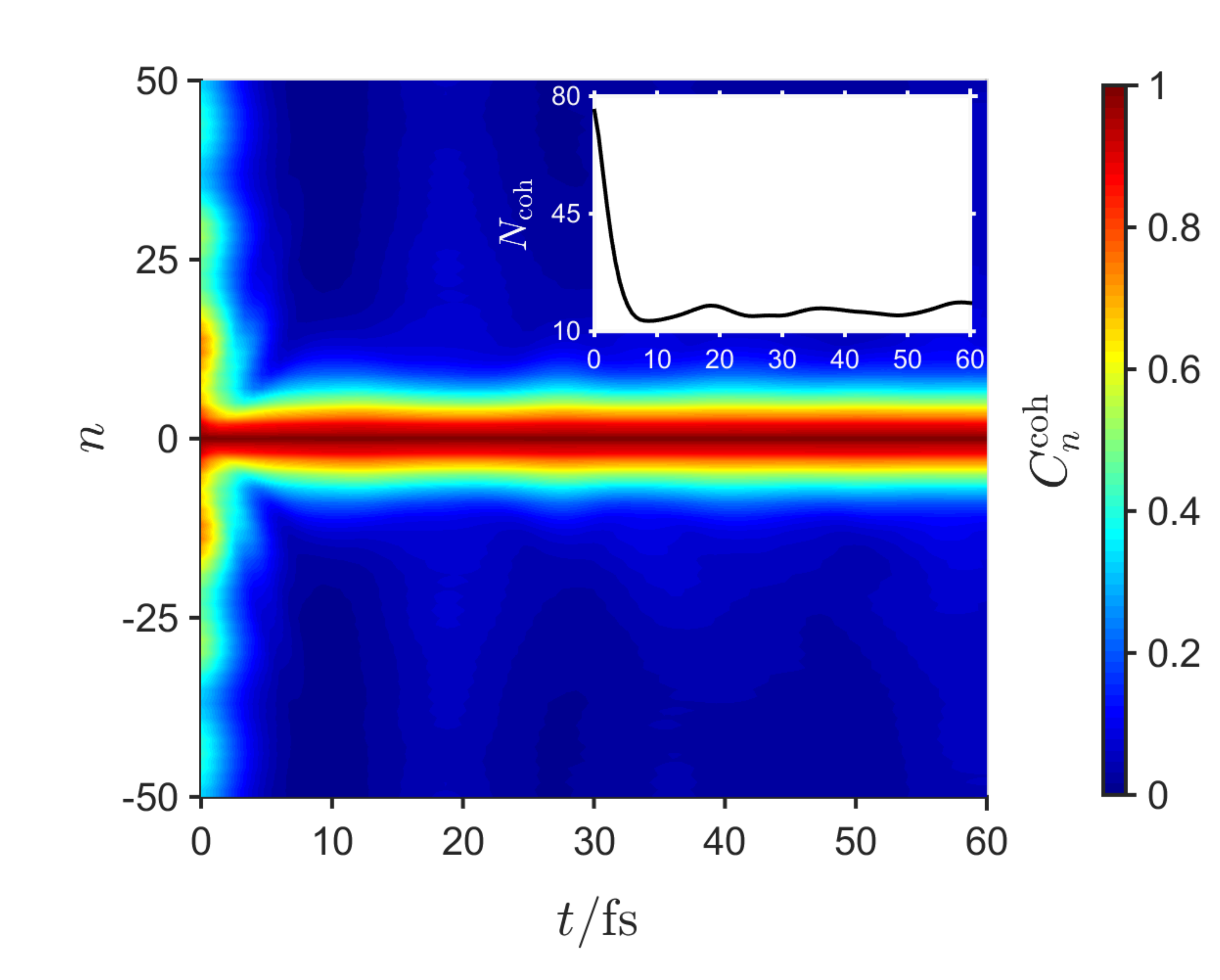}
\caption{\label{fig:corr_ex_ex} The time dependence of the exciton coherence correlation function, $C_{n}^{\text{coh}}$, for an initial high energy QEES given in Fig.~\ref{fig:qees}. In addition, the time dependence of the associated coherence number, $N_{\text{coh}}$, is given in the inset figure.}
\end{figure}

Figure~\ref{fig:corr_ex_ex} shows the time dependence of the exciton coherence correlation function for an initial high energy QEES given in Fig.~\ref{fig:qees}. The correlation function rapidly localizes, showing that within our model, decoherence of the exciton occurs on an ultra-fast time scale. This time scale is more evident from the time dependence of the exciton coherence number, $N_{\text{coh}}$:
\begin{equation}
\label{eq:coh_number}
N_{\text{coh}}=\sum_{n}C_{n}^{\text{coh}}
\end{equation}
which corresponds to the average number of moieties over which exciton coherences persist and is given in the inset of Fig.~\ref{fig:corr_ex_ex}. Indeed, we find that the coherence number reaches its equilibrium value within ${\sim}10$~fs. As for exciton-polaron formation, the short time scale associated with this decoherence leads to the associated dynamics being largely independent of the external dissipation and depending solely on the Frenkel-Holstein parameters.

The exciton decoherence mechanism present within our model can be elucidated by considering the following state representation for the Frenkel-Holstein model:
\begin{equation}
\label{eq:ket_examp}
\ket{\psi}=\sum_{n}\psi_{n}\ket{n}\ket{V_{n}}
\end{equation}
Within this representation, $\psi_{n}$ corresponds to the probability amplitude for the Frenkel exciton residing on site/moitey $n$, given by $\ket{n}$, while the ket $\ket{V_{n}}$ corresponds to the state of the $L$ harmonic oscillators associated with the exciton on site~$n$. Using this state representation, the exciton coherence correlation function takes the form:
\begin{equation}
\label{eq:corr_ex_ex_examp}
C_{n}^{\text{coh}}=\sum_{m}\left|\psi_{m}\psi_{m+n}^{*}\braket{V_{m+n}|V_{m}}\right|
\end{equation}
This expression contains two terms. The first term, $\psi_{m}\psi_{m+n}^{*}$, corresponds to the exciton wavefunction overlap between the two sites/moieties. As shown in Sec.~\ref{sect:localization}, this term remains essentially stationary over the time scale of exciton decoherence. The second term, $\braket{V_{m+n}|V_{m}}$, corresponds to the overlap of the vibrational states associated with the exciton being on sites/moieties $m+n$ and $m$. In Sec.~\ref{sect:polaron}, we saw that during the exciton relaxation dynamics, exciton-polaron formation occurs on an ultra-fast time scale, leading to the nuclear displacements of the polymer becoming locally correlated to the exciton. This means that during the dynamics, the vibrational state $\ket{V_{m}}$ will only have nuclear displacements spatially close to moiety/site~$m$ (where the exciton resides), thus causing the vibrational overlap $\braket{V_{m+n}|V_{m}}$ to decrease with increasing site separation $n$. It is this local nature of the vibrational overlap that accounts for the local nature of the exciton coherence correlation function in Fig.~\ref{fig:corr_ex_ex}. Exciton-polaron formation is therefore the mechanism by which the exciton decoheres, which is consistent with decoherence mechanisms found in other electron-nuclear coupled systems.\cite{overlap,overlap1,overlap2}

Knowledge of the exciton decoherence mechanism now gives insight into the dependence of the associated time scale on the Frenkel-Holstein parameters. Based on this mechanism, we would expect the exciton decoherence time scale to be identical to that for exciton-polaron formation, depending solely on $\hbar\omega$ and with decoherence occurring within half a vibrational time period. While this is consistent with our results, we find that the exciton decoherence time scale additionally depends on the exciton-phonon coupling strength, $A$, with the time scale decreasing with increasing $A$. This arises, because increasing $A$ leads to an increase in the nuclear displacements associated with the exciton-polaron, which results in a more rapid decay in the vibrational overlaps that lead to decoherence.

\subsection{Exciton Density Localization}
\label{sect:localization}
In Sec.~\ref{sect:decohere}, we saw that rapid exciton decoherence occurs during the relaxation dynamics, driven by the decay in the overlap of the vibrational states associated with exciton-polaron formation. As exciton decoherence arises from the vibrational configuration of the system, the exciton density can in principle be delocalized along the polymer chain, even if the exciton coherences between different sites/moieties are short ranged. Therefore, if the system does relax to the low energy VRSs, the exciton density must also localize during the time evolution.

\begin{figure}
\includegraphics[width=8.5cm]{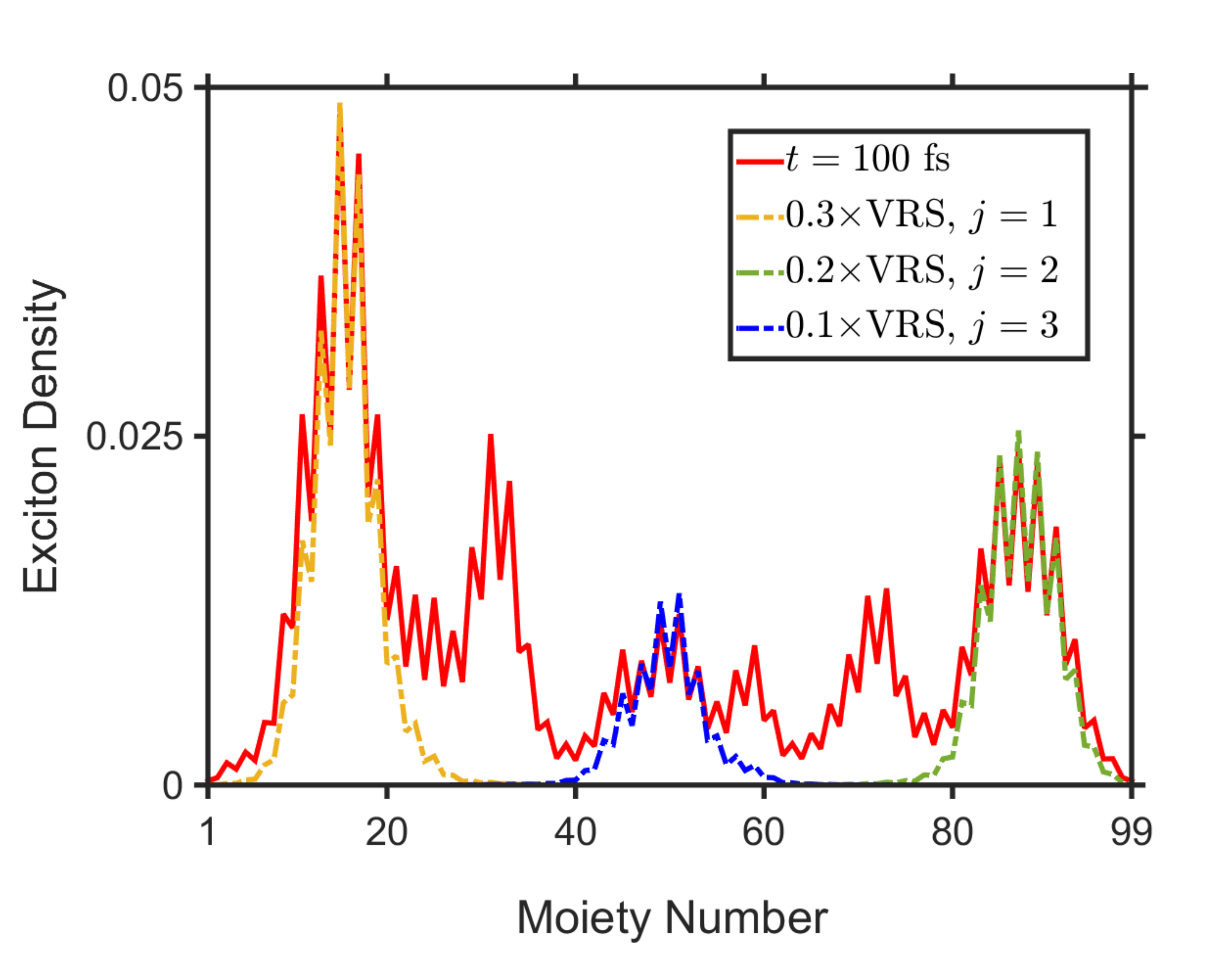}
\caption{\label{fig:exciton_dens} The exciton density at $t=100$~fs, given by the solid curve, for an initial QEES given in Fig.~\ref{fig:qees}. Also present are the scaled exciton densities for the three VRSs of the polymer chain, given by the dotted curves.}
\end{figure}
In Fig.~\ref{fig:exciton_dens}, the solid curve corresponds to the exciton density at time $t=100$~fs for an initial high energy QEES given in Fig.~\ref{fig:qees}. Also shown are the scaled exciton densities for the three VRSs of the polymer chain, given by the dotted curves. Three of the peaks in the exciton density at $t=100$~fs match those corresponding to the VRSs, suggesting that the system is relaxing into these low energy states during the time evolution. The presence of additional peaks in the time evolved exciton density are probably a result of the system having not reached the steady state at $t=100$~fs, but could also signify that the steady state solutions of our Lindblad master equation do not correspond to the VRSs. As we are primarily interested in understanding the short time dynamics in this paper, we will pursue this potential issue further in a subsequent paper, where the steady state solutions of the Lindblad master equation will be investigated.

While the time evolved exciton density contains peaks corresponding to the VRSs of the polymer chain, the solid curve in Fig.~\ref{fig:exciton_dens} seems to show that the exciton density remains quasi-delocalized during the dynamics. However, as the system is described by a mixed state density matrix, the associated observables physically correspond to an ensemble average over many different environment configurations (or alternatively, over many quantum trajectories). This means that it is not \emph{a priori} obvious whether the non-local nature of the ensemble averaged exciton density is symptomatic of an actual absence of exciton density localization in our model
or whether it is simply a consequence of ensemble averaging over
exciton density that has localized onto different chromophores. To distinguish between these two possibilities requires a correlation function that measures the spatial extent of the exciton for a single environment configuration, such as:\cite{local,local1}
\begin{equation}
\label{eq:ex_loc}
C_{n}^{\text{loc}}=\frac{\sum_{m}\left|\braket{m,0|\hat{\rho}|m+n,0}\right|}{\sum_{m}\left|\braket{m,0|\hat{\rho}|m,0}\right|}
\end{equation}
where the scaling factor is chosen so that $C_{0}^{\text{loc}}=1$. In this definition, the ket $\ket{m,0}$ corresponds to the exciton being on site/moiety~$m$, while all the $L$ harmonic oscillators are in their ground state, signified by the `0' index.

The physical interpretation of this exciton localization correlation function can be understood by considering the state representation for the Frenkel-Holstein model, given in Eq.~(\ref{eq:ket_examp}). Using this state representation, the correlation function takes the form:
\begin{equation}
\label{eq:ex_loc_examp}
C_{n}^{\text{loc}}=\frac{\sum_{m}\left|\psi_{m}\psi_{m+n}^{*}\braket{V_{m+n}|0}\braket{0|V_{m}}\right|}{\sum_{m}\left|\psi_{m}\right|^{2}\left|\braket{V_{m}|0}\right|^{2}}
\end{equation}
where $\psi_{m}$ is the probability amplitude for the exciton being on moiety/site~$m$, $\ket{V_{m}}$ corresponds to the state of the nuclear degrees of freedom when the exciton resides on site~$m$ and $\ket{0}$ corresponds to the ground state of the $L$ harmonic oscillators in the system. To a good approximation, the quantity $\braket{V_{m}|0}$ will be independent of the exciton site index $m$. This is because the spatial distribution of the oscillator displacements around the exciton depend on the exciton-phonon correlation length, which will be largely independent of the exciton site index. Applying this approximation to Eq.~(\ref{eq:ex_loc_examp}) leads to the following simplified form for the exciton localization correlation function:
\begin{equation}
\label{eq:ex_loc_examp2}
C_{n}^{\text{loc}}\approx\sum_{m}\left|\psi_{m}\psi_{m+n}^{*}\right|
\end{equation}
This expression shows that the correlation function can be regarded as giving the average magnitude of the exciton wavefunction overlap between moieties/sites a distance $n$ apart and therefore gives a measure of the spatial extent of the exciton on a polymer chain.

\begin{figure}
\includegraphics[width=8.5cm]{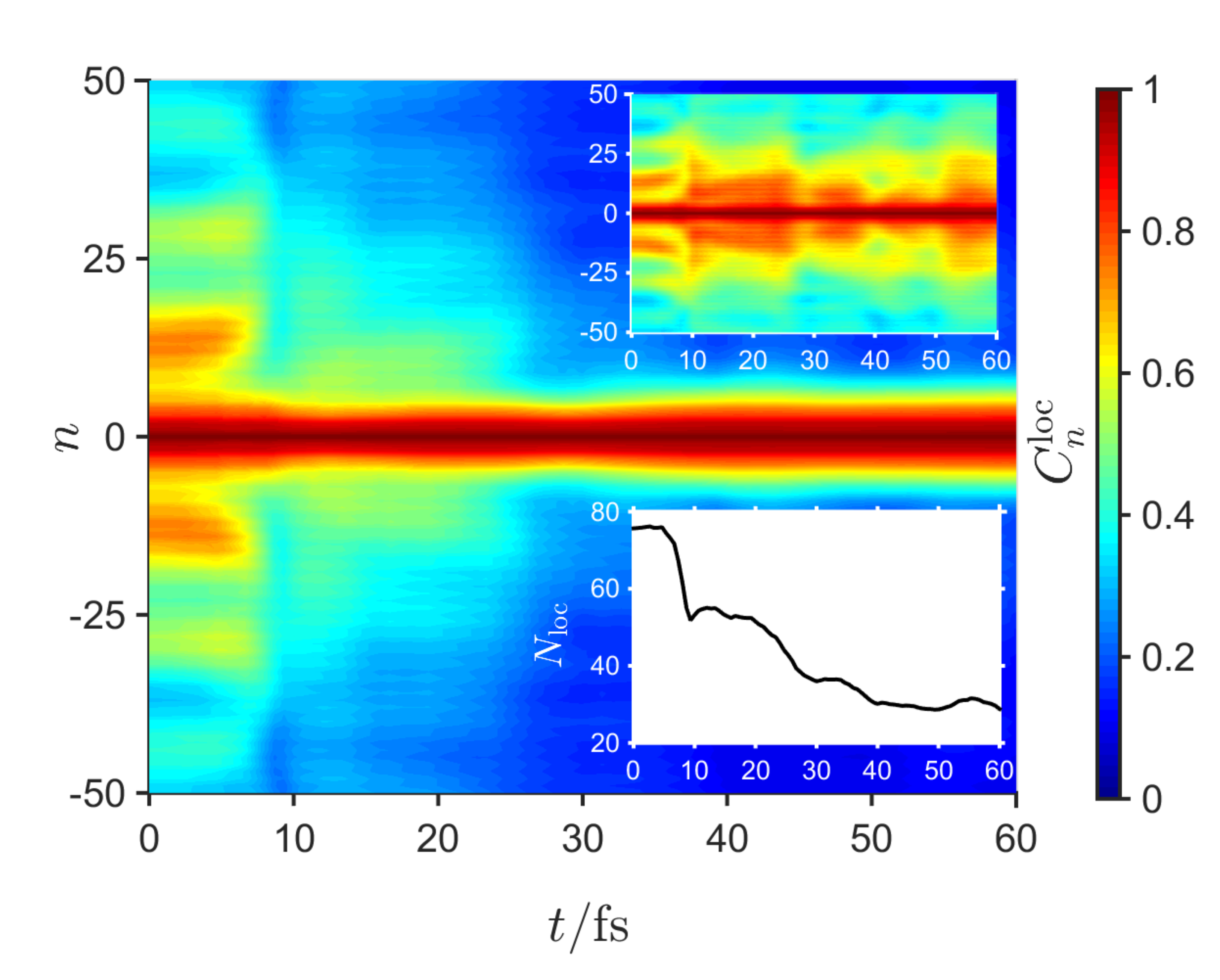}
\caption{\label{fig:corr_loc} The time dependence of the exciton localization correlation function, $C_{n}^{\text{loc}}$, for an initial high energy QEES given in Fig.~\ref{fig:qees}. The main figure corresponds to the time evolution with $\tilde{\gamma}=0.033$, while the upper inset figure corresponds to the time evolution without external dissipation. In addition, the time dependence of the associated exciton localization number, $N_{\text{loc}}$, with $\tilde{\gamma}=0.033$ is given in the lower inset figure.}
\end{figure}
Figure~\ref{fig:corr_loc} shows the time dependence of the exciton localization correlation function for an initial high energy QEES given in Fig.~\ref{fig:qees}. The main figure corresponds to the time evolution with the standard external dissipation parameter of $\tilde{\gamma}=0.033$, while the upper inset corresponds to the time evolution without external dissipation to the environment. When external dissipation is present, this correlation function does localize, confirming that exciton density localization is present within our model of the relaxation dynamics. However, the time scale associated with this process appears substantially different from the ultra-fast exciton-polaron formation and exciton decoherence time scales studied previously.
Indeed, the exciton localization correlation function remains essentially static for initial times on the order of the ultra-fast time scale (${\sim}10$~fs) with the function only starting to localize at longer times. This time scale is more evident from the time dependence of the exciton localization number, $N_{\text{loc}}$:\cite{local1}
\begin{equation}
\label{eq:loc_number}
N_{\text{loc}}=\sum_{n}C_{n}^{\text{loc}}
\end{equation}
which corresponds to the average number of moieties over which the exciton wavefunction overlap remains non zero and is given in the lower inset of Fig.~\ref{fig:corr_loc}. The exciton localization number continues to decrease for $t>60$~fs, illustrating the much longer time scale associated with this process. This suggests that the exciton density localization is driven by the external dissipation to the environment. Further confirmation of this is given in the upper inset of Fig.~\ref{fig:corr_loc}, which shows that without external dissipation, the correlation function remains delocalized during the time evolution.

\begin{figure}
\includegraphics[width=8.5cm]{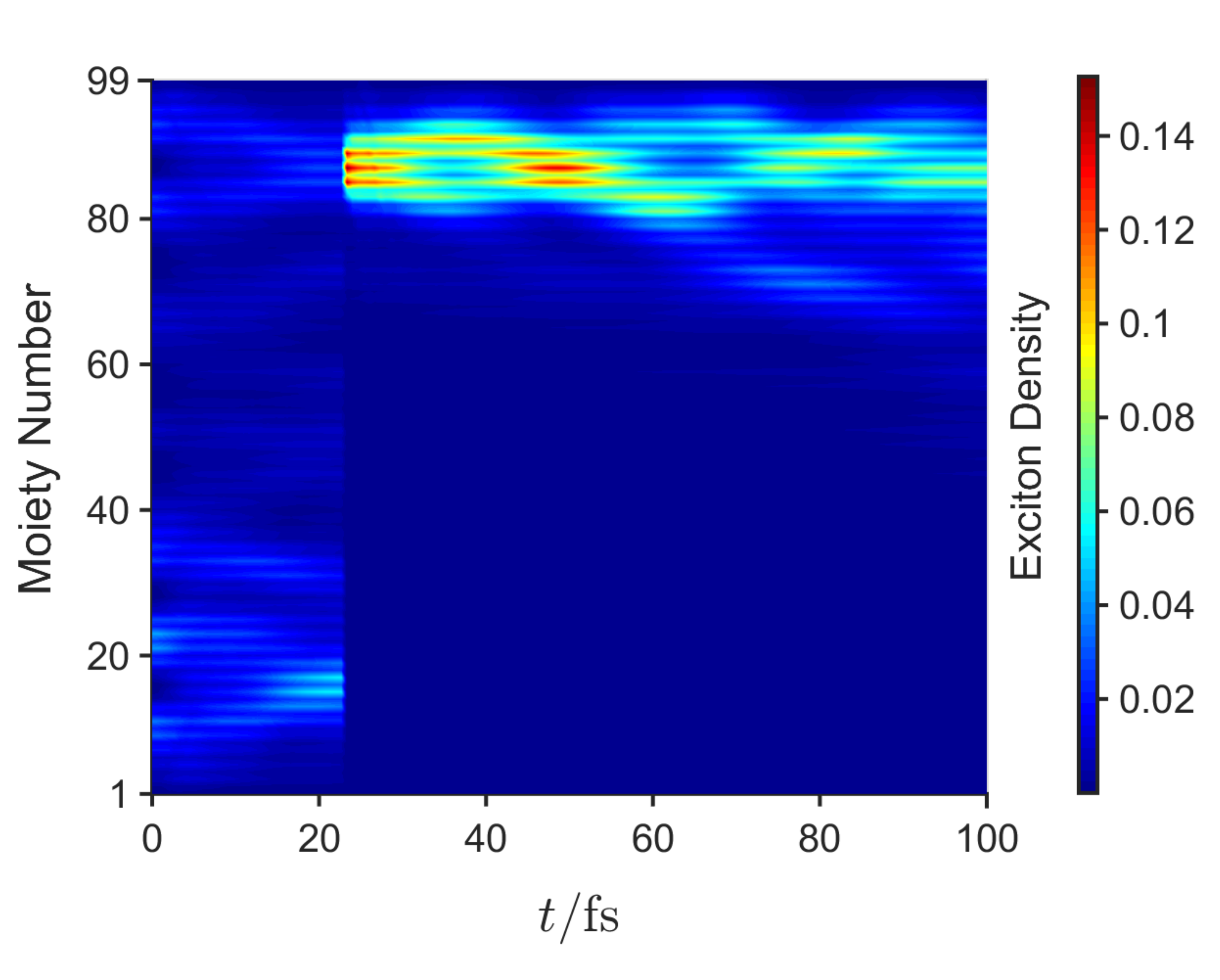}	\caption{\label{fig:trajectory} The time dependence of the exciton density for a single trajectory of the quantum jump trajectory method. 
The discontinuity in the density at ca.\ 20~fs is a `quantum jump' caused by the stochastic application of a Lindblad jump operator.
The dynamics were performed for an initial high energy QEES given in Fig.~\ref{fig:qees}, with $\tilde{\gamma}=0.033$.}
\end{figure}
The mechanism by which external dissipation drives this exciton density localization can be understood by investigating the time evolution of individual trajectories within the quantum jump trajectory method. The quantum jump trajectory method is a way of simulating the time evolution described by a Lindblad master equation and is outlined briefly in Sec.~\ref{sect:numerical}, with a more detailed summary given in Appendix~\ref{sect:jump}. Figure~\ref{fig:trajectory} shows the time dependence of the exciton density for a typical trajectory associated with this technique, where the system is initially in a high energy QEES given in Fig.~\ref{fig:qees}. For $t<20$~fs, the dynamics correspond to time evolution under the effective Hamiltonian, $\hat{H}_{\text{eff}}$ given in Eq.~(\ref{eq:heff}), where the figure shows that during this evolution, the exciton density remains quasi-delocalized along the polymer chain. The discontinuity in the exciton density at $t\sim20$~fs corresponds to a `quantum jump', in which one of the harmonic oscillator destruction operators, $\hat{b}_{n}$, is randomly applied to the system. The effect of this `quantum jump' process is to cause an immediate localization of the exciton density, as seen in Fig.~\ref{fig:trajectory}. \footnote{Therefore, it is the lack of these `quantum jump' processes within the Ehrenfest dynamics approximation that lead to the unphysical bifurcation of the exciton density found in previous work.\cite{ehren1}} This arises because only states of the harmonic oscillator on site~$n$ with a non zero displacement remain after application of the operator, $\hat{b}_{n}$. As these nuclear displacements are locally correlated to the exciton through exciton-polaron formation, application of $\hat{b}_{n}$ therefore leads to a state with the exciton localized around site~$n$. Physically this process corresponds to `wavefunction collapse', where local environment interactions with the internal phonon degrees of freedom act as a quantum measurement. Hence, it is the position of these local environment interactions that determines the particular chromophore that the exciton density relaxes onto during the time evolution.

\subsection{LEGSs Dynamics}
In the preceding analysis, the main focus has been on understanding the short time exciton relaxation dynamics of initial QEESs. However, the features of the dynamics, as well as their associated time scales, are largely identical for initial LEGSs, with a few important differences. For QEESs, we saw that the system evolves to become a mixed state combination of the various VRSs on the polymer chain, as seen in the time evolved exciton density. In contrast, we find that an initial LEGS adiabatically relaxes almost entirely onto a single VRS, with the exciton density remaining on the same chromophore throughout the time evolution. This finding is in agreement with previous work on the relaxation dynamics of LEGSs.\cite{legs_relax,ehren1} Finally, as the spatial extent of LEGSs and VRSs are very similar (see Fig.~\ref{fig:dmrg}) the extent of exciton decoherence and exciton density localization during the time evolution is much less pronounced compared to that for an initial QEES, resulting in a smaller time dependence of the associated correlation functions.

\subsection{Time Resolved Fluorescence Anisotropy}
So far, we have outlined many features present in our model of the exciton relaxation dynamics in polymer systems, such as exciton-polaron formation, exciton decoherence and exciton density localization. We now consider whether such features can be seen in experimental observables used to study the dynamics.

One such experimental technique is the measurement of the time-dependent decay in the fluorescence anisotropy after photoexcitation. It is known experimentally that as the exciton relaxes, the polarization axis associated with the fluorescence rotates from that of the incident radiation, caused by a rotation of the transition dipole moment of the polymer. This can be quantified using the fluorescence anisotropy, $r$:\cite{fluor_book}
\begin{equation}
\label{eq:r}
r = \frac{I_{\parallel}-I_{\perp}}{I_{\parallel}+2I_{\perp}}
\end{equation}
where $I_{\parallel}$ and $I_{\perp}$ are the intensities of the fluorescence radiation polarised parallel and perpendicular to the incident radiation, respectively.

For an arbitrary state of a quantum system, $\ket{\psi}$, the fluorescence intensity polarised along the $x$ axis is related to the $x$ component of the transition dipole operator, $\hat{\mu}_{x}$, by:
\begin{equation}
\label{eq:intens_ket}
I_{x} \propto \sum_{v}\left|\braket{\psi|\hat{\mu}_{x}|0,v}\right|^{2}
\end{equation}
where $\ket{0,v}$ corresponds to the system in the ground electronic state, with the nuclear degrees of freedom in the state characterised by the quantum number $v$. In principle, the expression for $I_{x}$ also has an energy dependent term. However, as long as the variance of the energy associated with the system remains small during the time evolution, this term just becomes a multiplicative constant that can be neglected in Eq.~(\ref{eq:intens_ket}). The expression given in Eq.~(\ref{eq:intens_ket}) can be generalised for a mixed state density matrix, $\hat{\rho}$, as follows:\cite{fluor_examp}
\begin{align}
\label{eq:intens_den}
\begin{split}
I_{x} & \propto \sum_{v}\braket{0,v|\hat{\mu}_{x}\hat{\rho}\hat{\mu}_{x}|0,v} \\
& \propto\sum_{m,n}s_{m,x}s_{n,x}\braket{m|\hat{\rho}_{\text{ex}}|n}
\end{split}
\end{align}
where the final equation is obtained by using the expression for the transition dipole moment given in Eq.~(\ref{eq:dipole}). In this equation, $\hat{\rho}_{\text{ex}}$ corresponds to the exciton reduced density matrix, defined in Eq.~(\ref{eq:dens_ex}), while $\ket{n}$ is the ket corresponding to the exciton residing on site~$n$. In addition, $s_{n,x}$ is the $x$ component of the unit vector that points along the polymer axis at site~$n$, illustrated in Fig.~\ref{fig:conformation}. Inserting the required intensity components, given in Eq.~(\ref{eq:intens_den}), into Eq.~(\ref{eq:r}) allows the fluorescence anisotropy to be obtained for a specific polymer conformation. The fluorescence anisotropy can then be averaged over several different polymer conformations and initial exciton states using the following expression:\cite{fluor_book}
\begin{equation}
\label{eq:intens_av}
\braket{r\left(t\right)} = 0.4\times\frac{\sum_{i}I_{i}\left(t\right)r_{i}\left(t\right)}{\sum_{i}I_{i}\left(t\right)}
\end{equation}
where $I_{i}\left(t\right)$ is the total fluorescence intensity and $r_{i}\left(t\right)$ is the fluorescence anisotropy, both associated with conformation $i$ at time $t$. The factor of 0.4 is included on the assumption that the polymers are oriented uniformly in the bulk material.\cite{fluor_book}

\begin{figure}
\includegraphics[width=8.5cm]{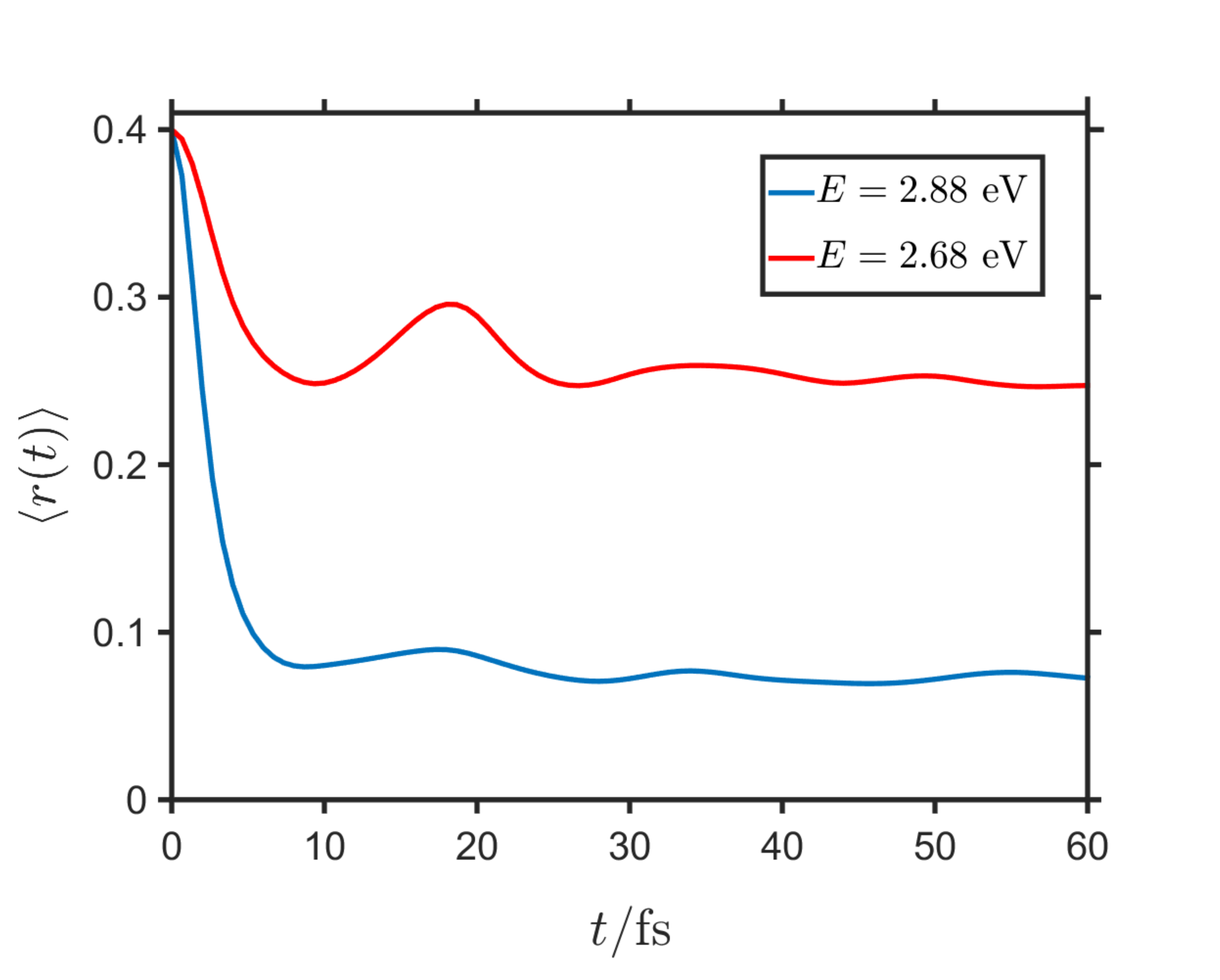}
\caption{\label{fig:r2} The time dependence of the fluorescence anisotropy, $\braket{r\left(t\right)}$, for two initial Frenkel exciton energies. The red curve corresponds to an initial Frenkel exciton energy of $2.68$~eV, while the blue curve corresponds to an initial Frenkel exciton energy of $2.88$~eV}
\end{figure}
Figure~\ref{fig:r2} shows the calculated time dependence of the fluorescence anisotropy for two different initial Frenkel exciton energies. The mechanism for the decay of the fluorescence anisotropy can be understood by comparing the expression for the components of the fluorescence intensity given in Eq.~(\ref{eq:intens_den}) with the expression for the exciton coherence correlation function given in Eq.~(\ref{eq:ex-ex-corr}). While the fluorescence intensity components cannot be directly written in terms of the exciton coherence correlation function, since both quantities depend on the elements of the exciton reduced density matrix, the time dependence of the two quantities are related. Indeed, we find numerically that the time dependence of the fluorescence intensity components are dominated by the decay of the off diagonal elements of the exciton reduced density matrix, which also give rise to the localization of the exciton coherence correlation function. The decay in the fluorescence anisotropy can therefore be attributed to the decoherence of the exciton during the relaxation dynamics, induced by exciton-polaron formation. This is also consistent with the observations that the decay of the fluorescence anisotropy is independent of $\tilde{\gamma}$ and also occurs on a time scale equal to the exciton decoherence time scale from Fig.~\ref{fig:corr_ex_ex}.

The difference in the magnitude of the fluorescence anisotropy decay for the two initial Frenkel exciton energies can now be explained in terms of this exciton decoherence mechanism. For the lower initial Frenkel exciton energy of $2.68$~eV, Fig.~\ref{fig:optical} shows that a large fraction of the initial Frenkel exciton states are LEGSs. As the exciton is already localized onto a single chromophore in a LEGS, the extent of exciton decoherence during the relaxation dynamics is much reduced for an initial LEGS compared to an initial QEES. This therefore explains why the magnitude of the fluorescence anisotropy decay is smaller for the lower initial Frenkel exciton energy in Fig.~\ref{fig:r2}, consistent with experimental observations.\cite{fluor}

\section{Discussion and Conclusions}
\label{sect:conclusion}
In this paper, we have introduced a model for exciton relaxation dynamics in polymer systems. This model is based on describing the system in terms of a parameterized Frenkel-Holstein Hamiltonian, which explicitly includes exciton-phonon coupling. While this Hamiltonian has been previously used to describe exciton relaxation dynamics in polymer systems,\cite{ehren1} our approach differs as we treat the nuclear degrees of freedom quantum mechanically, with external dissipation of these modes included in a Lindblad master equation formalism. Our dynamics simulations are then performed numerically using the quantum jump trajectory and time evolving block decimation techniques.

We find that our initial results for the time evolution are physically sensible, with the total system energy decreasing as a function of time, confirming that the dissipation in our model is having the desired effect. The values of the parameters relevant to polymer systems naturally lead to a separation of time scales within the dynamics. The ultra-fast time scale depends on the Frenkel-Holstein parameters and physically corresponds to energy transfer between the exciton and nuclear degrees of freedom, while the longer time dynamics are driven by the external dissipation parameter, $\tilde{\gamma}$, corresponding to damping of these phonon modes by the environment. These time scales are also found to be relevant to several features of our dynamics model, such as exciton-polaron formation, exciton decoherence and exciton density localization, which are symptomatic of the system relaxing to the so-called vibrational relaxed states (VRSs). In general we find:
\begin{itemize}
\item Exciton-polaron formation occurs within half a vibrational time period of the internal phonon modes (i.e., the C-C bond oscillations), namely within 10~fs. We also showed that the time-evolved exciton-phonon correlation function is in good agreement with that for the VRSs, which is consistent with the dynamics relaxing to these states.
\item Exciton decoherence is driven by the decay in the vibrational overlaps associated with exciton-polaron formation. The time scale associated with this process is similar to that of exciton-polaron formation.
\item The exciton density of initially prepared quasi-extended states evolves to have peaks corresponding to the more localized VRSs, which is consistent with the system relaxing onto emissive chromophores. Exciton density localization, whose ultimate cause is disorder, is driven by the external dissipation through `wavefunction collapse'. Therefore, this process has a much longer associated time scale than for exciton-polaron formation and exciton decoherence.
\end{itemize}

We emphasize that one key result of this work is that exciton decoherence and exciton density localization, which are terms that are often used interchangeably in the literature, are shown to be different physical processes that occur on different time scales. While exciton decoherence occurs on an intrinsic ultra-fast time scale of ${\sim}10$~fs as a result of the vibrational overlaps associated with exciton-polaron formation, exciton density localization (or more precisely, localization of the exciton-polaron itself) \cite{polaron_holst,polaron_landau} occurs on a much longer extrinsic time scale that is driven by the dissipation to the environment through a `wavefunction collapse' process.

We now attempt to interpret experimental observations in the light of these simulations. As stated in the introduction, experimental studies of the exciton relaxation dynamics in PPV observe two sub picosecond time scales (${\lesssim}50$~fs and ${\sim}100$~fs). Fluorescence depolarization experiments on PPV observe that the initial anisotropy value is ${<}0.4$, suggesting that an ultra-fast relaxation process is occurring with a time scale shorter than the time resolution of the experiment. This is consistent with the anisotropy decay time scale in our model and therefore we assign this ultra-fast process to the exciton decoherence arising from coupling to the high frequency C-C bond oscillations.

We also noted, however, that our calculated fluorescence anisotropy decay is essentially independent of the dissipation time scale, $\gamma^{-1}$, and therefore our simulation cannot account for the time scale of ${\sim} 100$~fs observed in fluorescence depolarization experiments. As suggested in previous work,\cite{fluor2} we predict that this further decay in the anisotropy arises from coupling of the exciton to the low frequency torsional modes in the polymer, which have an oscillation period of ca.\ 100~fs.\cite{torsion,ehren3} Indeed, as these torsional modes behave classically, they would be expected to form self-localized Landau polarons\cite{polaron_holst,polaron_landau} and therefore cause further exciton decoherence in addition to that arising from coupling to the high frequency C-C bond oscillations. However, because classical normal modes give rise to a diverging exciton-phonon correlation length, the exciton decoherence and exciton density localization processes will occur via different mechanisms than for the high frequency modes. As most spectroscopic quantities depend on the exciton reduced density matrix, it is likely that the two time scales seen in these experiments arise from exciton decoherence through coupling of the exciton to these two different normal modes.

The role of the low frequency torsional modes on exciton decoherence and spectroscopic measurements will be the subject of future work. Also outstanding is the question as to what are the experimental observables associated with exciton density localization.\footnote{Our simulations indicate that the fluorescence anisotropy defined by emission into the zeroth vibronic peak is qualitatively affected by exciton density localization.} We intend to compute the coherent electronic 2D spectra and three-pulse photon-echo spectra in an attempt to address this question. Finally, we are in the process of preparing a paper that describes the steady state solutions of the Lindblad master equation introduced here, to ascertain whether our dissipation model does cause the system to relax into the low energy VRSs, as physically expected.

\section{Acknowledgements}
This work was performed using the Tensor Network Theory Library, Beta Version 1.2.1 (2016), S. Al-Assam, S. R. Clark, D. Jaksch and TNT Development team, www.tensornetworktheory.org.
JRM would like to thank the following for financial support: the EPSRC Centre for Doctoral Training, Theory and Modelling in Chemical Sciences, under grant EP/L015722/1, as well as University College Oxford, through the Radcliffe Scholarship. 
The authors would like to acknowledge the use of the University of Oxford Advanced Research Computing (ARC) facility in carrying out this work (http://dx.doi.org/10.5281/zenodo.22558). We also thank S. R. Clark for helpful discussions in the early stages of this work.

\bibliography{paper_relax}

\appendix

\section{The Quantum Jump Trajectory Method}
\label{sect:jump}
The quantum jump trajectory method is a technique that allows the numerical solution of a Lindblad master equation, such as Eq.~(\ref{eq:master}), to be obtained using time-dependent Schr\"odinger equation based methods. More specifically, observables associated with the time-dependent density matrix solution of a Lindblad master equation are obtained by averaging over many so-called quantum jump trajectories. The time evolution of a single quantum jump trajectory is obtained using the following procedure:

\begin{itemize}
\item \textit{Step 1:} We first spilt up the total time evolution of a single trajectory into several time steps of length~$\delta t$. We next define an effective Hamiltonian for the system as follows:\cite{jump}
\begin{equation}
\label{eq:effective_ham}
\hat{H}_{\text{eff}} = \hat{H} - \frac{i\gamma}{2}\sum_{n}\hat{b}_{n}^{\dagger}\hat{b}_{n}
\end{equation}
where $\hat{H}$ is the system Hamiltonian, $\gamma$ is the Lindblad master equation dissipation parameter and $\hat{b}_{n}$ is the Lindblad operator for site~$n$. Evolving this single quantum trajectory, initially in state $\ket{\psi(t)}$, under the effective Hamiltonian gives us a trial state at time $t+\delta t$:
\begin{equation}
\label{eq:trial}
\ket{\psi_{\text{trial}}(t+\delta t)} = e^{-i\hat{H}_\text{eff}\delta t}\ket{\psi(t)}
\end{equation}

\item \textit{Step 2:} As the effective Hamiltonian in Eq.~(\ref{eq:effective_ham}) is non-Hermitian, the time evolution in Eq.~(\ref{eq:trial}) does not conserve the quantum state's norm:
\begin{equation}
\label{eq:norm}
\braket{\psi_{\text{trial}}(t+\delta t)|\psi_{\text{trial}}(t+\delta t)} = 1-\delta p
\end{equation}
where $\delta p$ is the amount the norm has decayed over the time step $\delta t$. Using Eq.~(\ref{eq:norm}), the state of the single quantum trajectory at time $t+\delta t$ is determined probabilistically as follows:
\begin{itemize}
\item[a)] With probability $1-\delta p$, the state at time $t+\delta t$ is:
\begin{equation}
\label{eq:timestep1}
\ket{\psi(t+\delta t)} = \frac{e^{-i\hat{H}_\text{eff}\delta t}\ket{\psi(t)}}{\sqrt{1-\delta p}}
\end{equation}

\item[b)] With probability $\delta p$, the state at time $t+\delta t$ is:
\begin{equation}
\label{eq:timestep2}
\ket{\psi(t+\delta t)} =\frac{\hat{b}_{n}\ket{\psi(t)}}{\sqrt{\braket{\psi(t)|\hat{b}_{n}^{\dagger}\hat{b}_{n}|\psi(t)}}}
\end{equation}
\end{itemize}
where $\hat{b}_{n}$ is the Lindblad operator for site~$n$. The site at which the Lindblad operator is applied in Eq.~(\ref{eq:timestep2}) is again determined probabilistically, where the probability associated with applying the Lindblad operator at site~$n$ is given by:
\begin{equation}
P_{n} = \frac{\braket{\psi(t)|\hat{b}_{n}^{\dagger}\hat{b}_{n}|\psi(t)}}{\sum_{m}\braket{\psi(t)|\hat{b}_{m}^{\dagger}\hat{b}_{m}|\psi(t)}}
\end{equation}
\end{itemize}
Hence, we now need a technique by which to apply the time step evolution operator, $\text{exp}[-i\hat{H}_\text{eff}\delta t]$, for the system of interest. This is achieved for the Frenkel-Holstein model in this work using the TEBD technique, which is introduced in Appendix~\ref{sect:tebd}.

\section{Time Evolving Block Decimation}
\label{sect:tebd}
\subsection{Matrix Product States}
It has been found that a practical quantum state representation for one-dimensional systems involves using matrix product states (MPS). The MPS representation of a generic quantum state is as follows:\cite{mps}
\begin{equation}
\label{eq:mps}
\ket{\psi} = \sum_{\sigma_{1},..,\sigma_{L}}\sum_{\alpha_{1},..,\alpha_{L-1}}A^{\sigma_{1}}_{\alpha_{1}}A^{\sigma_{2}}_{\alpha_{1},\alpha_{2}}...A^{\sigma_{L}}_{\alpha_{L-1}}\ket{\sigma_{1}}\otimes....\otimes\ket{\sigma_{L}}
\end{equation}
where $\sigma_{n}$ are known as the physical indices, which correspond to the quantum numbers of the site basis, $\ket{\sigma_{n}}$, and $\alpha_{n}$ are known as the internal indices. For systems made up of $L$ sites, with a physical index dimension $d$, an arbitrary quantum state can be exactly represented by a MPS with an internal index dimension of $d^{L/2}$. However, most states can be accurately represented with a much smaller internal index dimension, where increased internal index dimensions are required to represent quantum states with a greater amount of entanglement between sites.

\begin{figure}
\includegraphics[width=7cm]{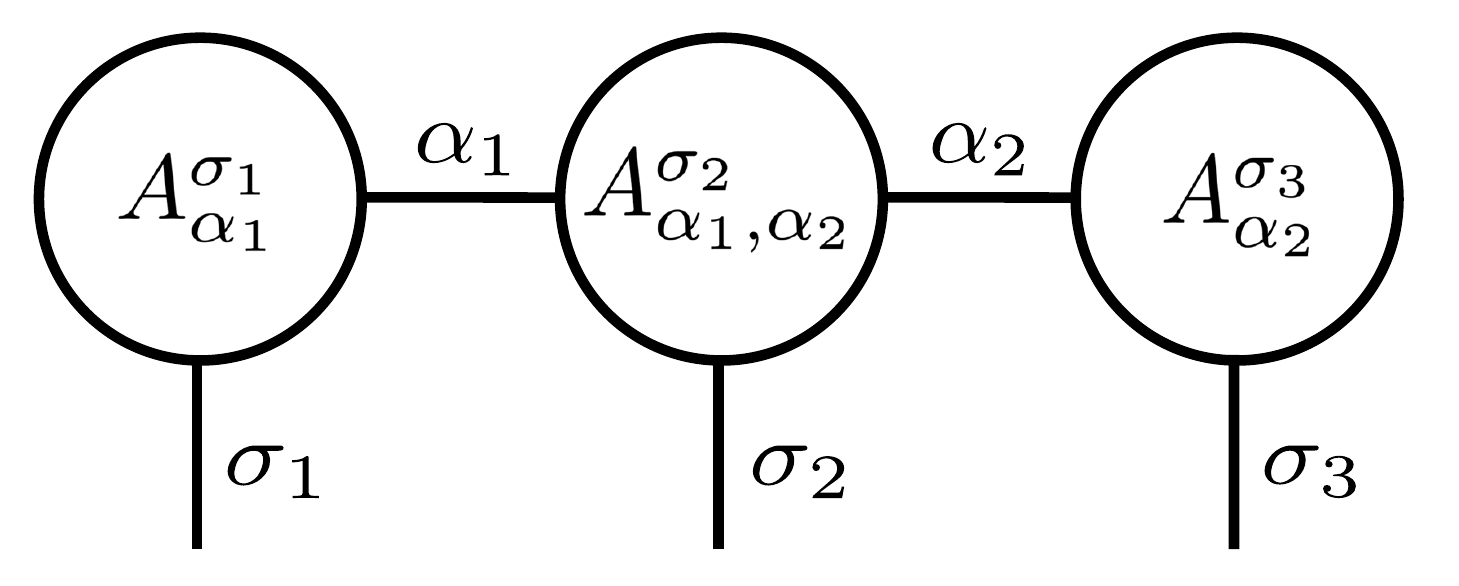}
\caption{\label{fig:mps} Tensor network diagram corresponding to a three site MPS.}
\end{figure}

MPSs can be helpfully represented and manipulated using tensor network diagrams.\cite{tnd} Figure~\ref{fig:mps} shows the tensor network diagram for a three site MPS. In these diagrams, the circles represent the tensors in the MPS, $A^{\sigma_{n}}_{\alpha_{n},\alpha_{n+1}}$, while the lines correspond to their indices. If a line is `closed', then the index it represents is implicitly summed over.
\subsection{Trotter Decomposition}
In order to obtain the quantum dynamics for a system, the evolution operator for a time step, $\text{exp}[-i\hat{H} \delta t]$, must first be computed. For systems that have a large Hilbert space, this is not possible. Within the TEBD technique, this issue is circumvented by using the Trotter decomposition to expand the evolution operator into several smaller and more manageable terms.

For Hamiltonians that contain solely on-site and nearest neighbor terms, it can be shown that the full Hamiltonian can be written as the following sum:
\begin{equation}
\label{eq:ham_bond}
\hat{H} = \sum_{n=1}^{L-1}\hat{H}_{n,n+1}
\end{equation}
where $\hat{H}_{n,n+1}$ is the Hamiltonian for the ``bond'' linking sites $n$ and $n+1$ and $L$ is the total number of sites in the system. Using Eq.~(\ref{eq:ham_bond}), the Trotter decomposition for the evolution operator is given by:\cite{trotter}
\begin{equation}
\label{eq:trotter}
e^{-i\hat{H}\delta t} \simeq e^{-\tfrac{i}{2}\hat{H}_{1,2}\delta t}e^{-\tfrac{i}{2}\hat{H}_{2,3}\delta t}...e^{-\tfrac{i}{2}\hat{H}_{2,3}\delta t}e^{-\tfrac{i}{2}\hat{H}_{1,2}\delta t} + \text{O}(\delta t^{3})
\end{equation}
The Trotter decomposition given in Eq.~(\ref{eq:trotter}) is not exact, as the Hamiltonian terms $\hat{H}_{n,n+1}$ in general do not commute with each other. However, the error associated with this Trotter decomposition is of $\text{O}(\delta t^{3})$, which becomes negligible if the time steps are chosen to be small.
\subsection{The Procedure}
\begin{figure*}
\includegraphics[width=15cm]{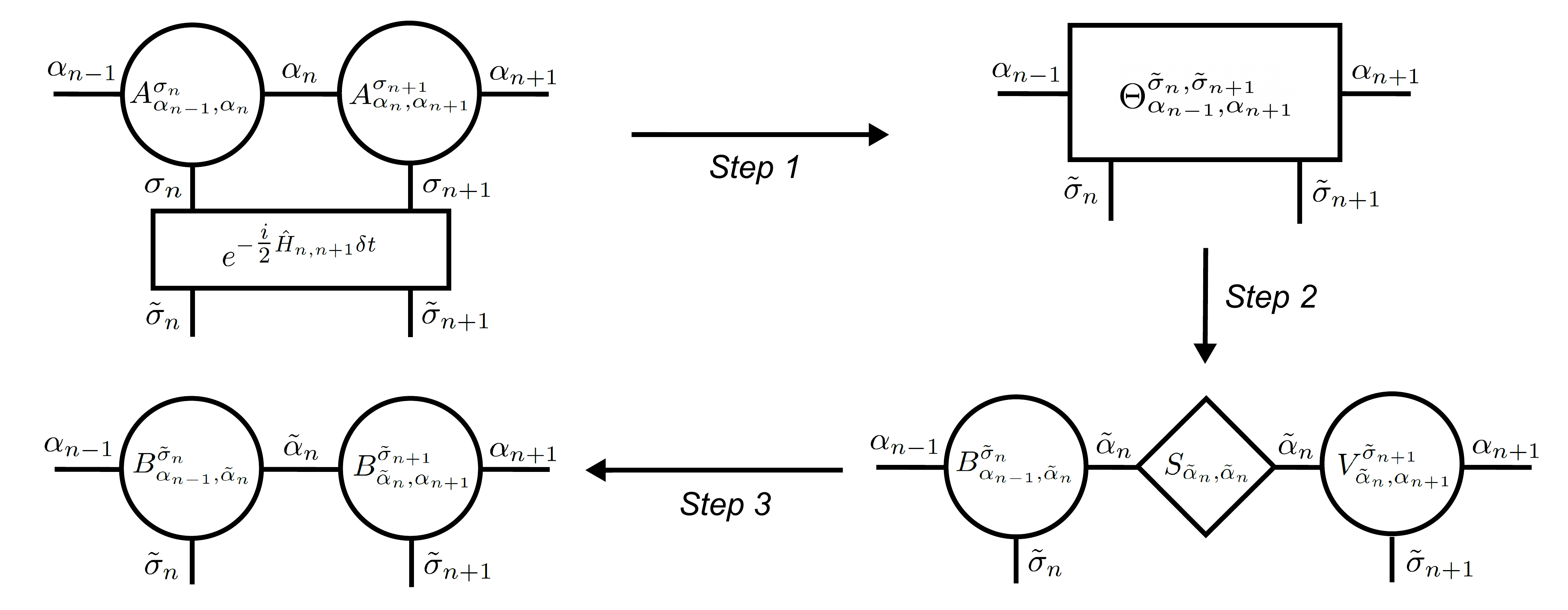}
\caption{\label{fig:tebd} Diagram outlining the TEBD procedure.}
\end{figure*}
The procedure for applying a single Trotter decomposition evolution operator onto a MPS is illustrated in Fig.~\ref{fig:tebd}, with the first tensor network diagram in this figure corresponding to the object $\text{exp}[-i\hat{H}_{n,n+1}\delta t/2]\ket{\psi}$. This figure illustrates the following steps:

\begin{itemize}
\item \textit{Step 1:} The closed indices $\alpha_{n}$, $\sigma_{n}$ and $\sigma_{n+1}$ are summed over to form the new tensor $\Theta_{\alpha_{n-1},\alpha_{n+1}}^{\tilde{\sigma}_{n},\tilde{\sigma}_{n+1}}$. From Fig.~\ref{fig:tebd}, we see that this new tensor is given by:\cite{tebd}
\begin{equation}
\label{eq:theta_def}
\Theta_{\alpha_{n-1},\alpha_{n+1}}^{\tilde{\sigma}_{n},\tilde{\sigma}_{n+1}} =\sum_{\sigma_{n},\sigma_{n+1},\alpha_{n}}U^{\tilde{\sigma}_{n},\tilde{\sigma}_{n+1}}_{\sigma_{n},\sigma_{n+1}}A_{\alpha_{n-1},\alpha_{n}}^{\sigma_{n}}A^{\sigma_{n+1}}_{\alpha_{n},\alpha_{n+1}}
\end{equation}
where
\begin{equation}
\label{eq:U_def}
U^{\tilde{\sigma}_{n},\tilde{\sigma}_{n+1}}_{\sigma_{n},\sigma_{n+1}} = \bra{\tilde{\sigma}_{n},\tilde{\sigma}_{n+1}}e^{-\tfrac{i}{2}\hat{H}_{n,n+1}\delta t}\ket{\sigma_{n},\sigma_{n+1}}
\end{equation}

\item \textit{Step 2:} A singular value decomposition is applied on the tensor $\Theta_{\alpha_{n-1},\alpha_{n+1}}^{\tilde{\sigma}_{n},\tilde{\sigma}_{n+1}}$ to obtain three new tensors, as well as a new internal index $\tilde{\alpha}_{n}$:\cite{tebd}
\begin{equation}
\label{eq:SVD}
\Theta_{\alpha_{n-1},\alpha_{n+1}}^{\tilde{\sigma}_{n},\tilde{\sigma}_{n+1}} = B^{\tilde{\sigma}_{n}}_{\alpha_{n-1},\tilde{\alpha}_{n}}S_{\tilde{\alpha}_{n},\tilde{\alpha}_{n}}V^{\tilde{\sigma}_{n+1}}_{\tilde{\alpha}_{n},\alpha_{n+1}}
\end{equation}
If the size of the original internal index $\alpha_{n}$ was $\chi$, then the new internal index $\tilde{\alpha}_{n}$ has grown to size $\chi d$ on completion of the procedure outlined above, where $d$ is the size of the site basis. To keep the internal indices constant in size, we truncate the $\tilde{\alpha}_{n}$ index by discarding the $\tilde{\alpha}_{n}$ `states' associated with the smallest values of $S_{\tilde{\alpha}_{n},\tilde{\alpha}_{n}}$.\cite{tebd} The error associated with this truncation procedure can be quantified by computing the sum of the discarded singular values, $S_{\tilde{\alpha}_{n},\tilde{\alpha}_{n}}$, which gives a bound on the error associated with the time evolved quantum state. It is this truncation procedure that keeps the total Hilbert space of the system finite and manageable throughout the dynamics.

\item \textit{Step 3:} The resulting tensor network diagram is transformed back into MPS form by creating the new tensor $B_{\tilde{\alpha}_{n},\alpha_{n+1}}^{\tilde{\sigma}_{n+1}}$:
\begin{equation}
B_{\tilde{\alpha}_{n},\alpha_{n+1}}^{\tilde{\sigma}_{n+1}} = S_{\tilde{\alpha}_{n},\tilde{\alpha}_{n}}V^{\tilde{\sigma}_{n+1}}_{\tilde{\alpha}_{n},\alpha_{n+1}}
\end{equation}
\end{itemize}
Hence to complete a dynamics time step within the TEBD algorithm, the same three step procedure outlined above is performed for every evolution operator in the Trotter decomposition, given in Eq.~(\ref{eq:trotter}).

Outlined above is the necessary framework for numerically obtaining the dynamics associated with the Lindblad master equation given in Eq.~(\ref{eq:master}). For our simulations of the exciton relaxation dynamics for PPV polymer chains containing 99 moieties/sites, the parameters used in the TEBD algorithm are given in Table~\ref{tab:param_tebd}. These were obtained as follows. The maximum number of phonons per site, $n_\text{phon}=2$, was chosen as it gives physically sensible results for the dynamics, while keeping the size of the site Hilbert space manageable. The simulations were also performed for $n_\text{phon}>2$, where we found that the results did not deviate significantly from the $n_\text{phon}=2$ calculations. Using dimensionless time, $\tilde{t}=\omega t$, the value of the time step, $\delta\tilde{t}$, was obtained by comparing the TEBD results for a four site Frenkel-Holstein model with the numerically exact dynamics obtained from exact diagonalization. As for the case of determining $n_{\text{phon}}$, we chose to use $\chi=150$ as a balance between obtaining accurate results for the quantum dynamics, as well as making the calculation not too computationally intensive. The error associated with this Hilbert space truncation (quantified from the sum of the discarded singular values, $S_{\tilde{\alpha}_{n},\tilde{\alpha}_{n}}$, obtained during the singular value decompositions within the TEBD algorithm) was found to be small for each time step during our dynamics simulations, suggesting that we can be confident that the Hilbert space truncation does not lead to a large error in our results for these parameter values.

\begin{table}[H]
\centering
{\renewcommand{\arraystretch}{1.2}
\begin{tabular}{|M{2cm}|M{2cm}|M{2cm}|M{2cm}|}
\hline
$\textbf{Parameter}$ & $\textbf{Value}$ & $\textbf{Parameter}$ & $\textbf{Value}$ \\
\hline
$\delta\tilde{t}$ & $5 \times 10^{-4}$ & $\chi$ & $150$ \\
$n_{\text{phon}}$ & 2 &  &   \\
\hline
\end{tabular}}
\caption{\label{tab:param_tebd} Parameters used in the time evolving block decimation technique (TEBD).}
\end{table}

\end{document}